# A Theory of Market Efficiency

Anup Rao
University of Washington
anuprao@cs.washington.edu

February 7, 2017

**Abstract**

We introduce a mathematical theory called *market connectivity* that gives concrete ways to both measure the efficiency of markets and find inefficiencies in large markets. The theory leads to new methods for testing the famous *efficient markets hypothesis* that do not suffer from the *joint-hypothesis* problem that has plagued past work. Our theory suggests metrics that can be used to compare the efficiency of one market with another, to find inefficiencies that may be profitable to exploit, and to evaluate the impact of policy and regulations on market efficiency.

A market's efficiency is tied to its ability to communicate information relevant to market participants. Market connectivity calculates the speed and reliability with which this communication is carried out via trade in the market. We model the market by a network called the *trade network*, which can be computed by recording transactions in the market over a fixed interval of time. The *nodes* of the network correspond to participants in the market. Every pair of nodes that trades in the market is connected by an *edge* that is weighted by the rate of trade, and associated with a vector that represents the type of item that is bought or sold.

We evaluate the ability of the market to communicate by considering how it deals with *shocks*. A shock is a change in the beliefs of market participants about the value of the products that they trade. We compute the effect of every *potential* significant shock on trade in the market. We give mathematical definitions for a few concepts:

- The *tension* and *energy* of the network are related concepts that measure the strength of the connections between sets of participants that trade similar items. They measure the amount of trade that is affected by significant shocks. They are high when there are many paths of high rate of trade that connect those with differing beliefs about the value of items. They are low when information from some large set of participants must take a long time to reach some other large set via trade.
- A *bottleneck* in the network is a small set of nodes that monopolizes an unusually large share of the trade in the network. The nodes in the bottleneck have an incentive to set prices incorrectly and interfere with the fair transmission of information in the market.

We give explicit mathematical definitions that capture these concepts and allow for quantitative measurements of market inefficiency.

# Contents

# 1 Introduction

Market-based economies have come to be the dominant system for the production and distribution of goods and services. The power of markets stems from their decentralized nature [Hay45]. A market is able to channel the effort of individuals in fruitful directions, even though no individual is aware of all the information necessary to decide how their effort ought to be expended. Complex decisions about how to react to scarcities, surpluses, and spikes in demand are handled without the intervention of a central entity. Over time, markets have proven themselves to be adept at taking advantage of economies of scale, and forming reliable and efficient distribution networks.

The forces of self-interest, competition, and the balancing of supply and demand all exert stabilizing influences on prices in the market. They suggest the concept of an *efficient market*, defined by Fama [Fam70, Fam91] as follows:

**Definition 1.** *An* efficient market *is one in which prices always fully reflect available information.*

It has long been theorized that financial markets are efficient, or close to being efficient. This is the well-known *efficient markets hypothesis* [Fam70, Fam91]. The efficient markets hypothesis is a seductive idea, an idea that is simultaneously deep and accessible. The hypothesis is supported by compelling intuition—if prices did not accurately reflect all information, then traders would have an opportunity to exploit the discrepancy to make a profit, which would lead to prices becoming more accurate. So, prices should not drift far from being correct. As Malkiel notes, "[...] [W]hen information arises, the news spreads very quickly and is incorporated into the prices of securities without delay." [Mal03, p. 59]

However, much is left open to interpretation in this reasoning and in Definition 1. What does it really mean that prices *fully reflect* information? What is the mechanism by which information spreads? If a market is not efficient, how can one quantify its inefficiency? Exactly what information is *available* to the market and what information is *not* available? These questions are not addressed by Definition 1 or the reasoning associated with the efficient markets hypothesis. Even if we knew *all* information, and we knew what information is available to all market participants, we might find it difficult to agree on what prices ought to be. If a tree falls in a forest and no one is around to hear it, should the price of wood change? What if a woodcutter hears the tree falling, how much should the price of wood change then? What if a carpenter hears the tree falling, how much should the price change then?

Still, the efficient markets hypothesis is extremely useful, and an extensive theory has been developed to test it. The definitional problems discussed above lead to a sticking point in this established theory, called the *joint-hypothesis problem.* As Fama explains it:

> [...] [M]arket efficiency per se is not testable. It must be tested jointly with some model of equilibrium, an asset-pricing model. [...] [W]e can only test whether information is properly reflected in prices in the context of a pricing model that defines the meaning of "properly." As a result, when we find anomalous evidence on the behavior of returns, the way it should be split between market inefficiency or a bad model of market equilibrium is ambiguous [Fam91, p. 1575].

The aim of this work is to develop a rigorous mathematical theory that can concretely quantify market inefficiencies and so replace Definition 1 with concepts that are testable. We wish to eliminate the joint-hypothesis problem and minimize the guesswork involved in deciding whether or not prices properly take into account information that they should. We seek a model that can be calibrated with data from the real world, and analyzed with computers to find and quantify inefficiencies.

Such a model could have far reaching consequences. Catastrophic recessions or dramatic swings in prices are less likely to occur in a market that is more efficient. An efficient market should experience a large change in prices only if market participants suddenly become aware of an unexpected *external* change in the world. A useful mathematical model could provide an important parameter that indicates the health of the market. Regulators and policy makers could use it to objectively evaluate their decisions. Firms could use it to identify inefficiencies that can be exploited for profit, helping to eliminate the inefficiencies.

## 1.1 Our Approach

We introduce a new mathematical theory of efficiency that we call *market connectivity*. Market efficiency is closely tied to a market's ability to communicate economic information via trade. So, market connectivity measures the speed and reliability with which the market is able to carry out this communication. Market participants do learn information from sources external to the market[1], but the theory focuses on the ability of the market itself to communicate information by trade; the advantage is that this can be measured with data. In our theory, a market that is able to disseminate information quickly and reliably is deemed efficient, and conversely, a market that cannot disseminate information quickly and reliably is deemed inefficient. Our concepts maintain the spirit of Definition 1, but avoid the joint-hypothesis problem, because we measure the ability of the market to communicate, rather than the accuracy of prices.

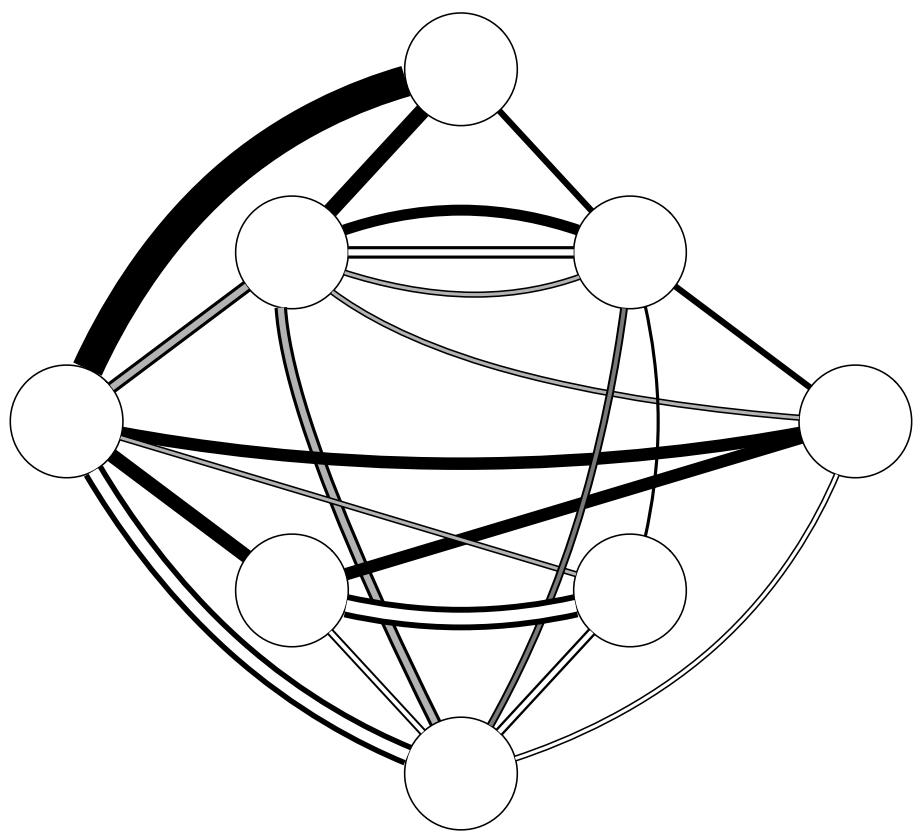

Figure 1: A trade network with 8 nodes, represented as circles. Edges connect nodes that trade. Edges are weighted by the rate of trade, and colored according to the type of item being traded.

We represent the market by a network called the *trade network*. Every participant in the market corresponds to a node of the network—every firm, individual, or other entity that buys and sells in the market is represented by a node. Every pair of nodes is connected by an edge weighted by the rate of trade between the corresponding participants, and associated with a vector that represents the type of the item being traded. The rate of trade can be estimated by computing the ratio of the total monetary value of the trade between the two participants during an interval of time (say a year), divided by the length of the interval. The vectors that represent the type of item can be computed from the distribution of trade in the item (more on this in Section 3). These type vectors encode the similarity of different items traded in the network. The trade network itself evolves with time[2] as traders discover new

[1] As we discuss at the end of this section, our theory is likely to account for all significant external channels of information as well, because such channels will lead to parallel channels of trade.

[2] The evolution of the trade network over time does encode some information about the efficiency of the market. It captures how quickly the market is able to form new trade connections in response to inefficiencies. We do not study this further here, but this seems to be a direction for future work that is worth pursuing. More on this in Section 5.

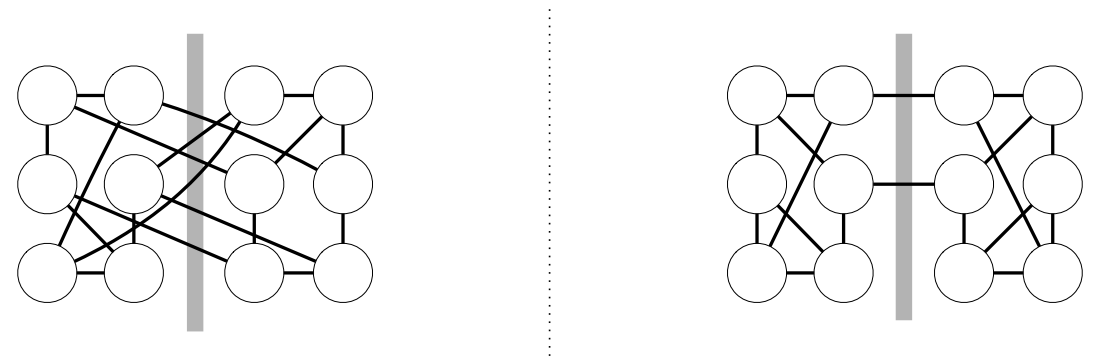

(a) Two trade networks with the same total volume of trade. Every node in both networks trades with 3 other nodes, and every edge has the same rate of trade, yet the network on the left is better connected. Information will take a long time to propagate across the gray divider in the network on the right.

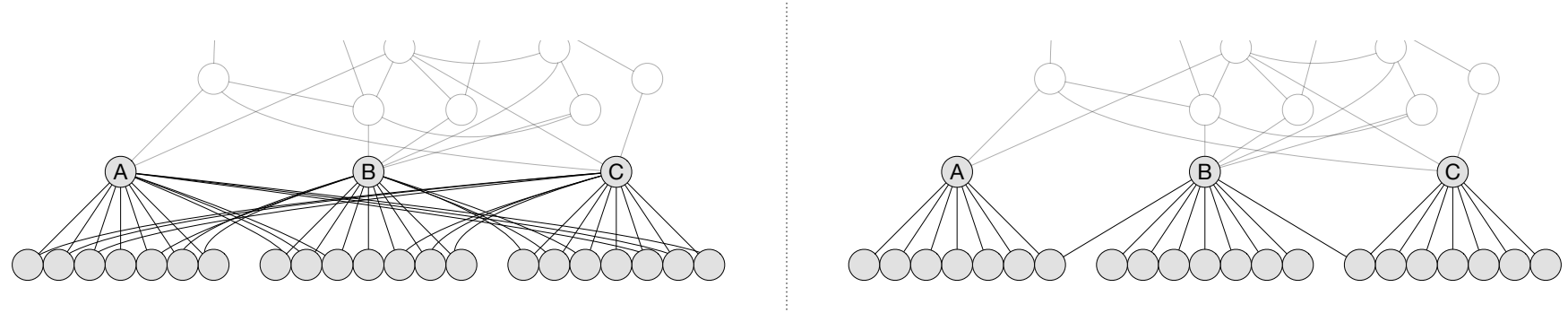

(b) Two trade networks where A,B,C provide a service to a set of customers. A,B,C have the same share of the market in both networks, but the network on the left is more efficient, because most of A's customers also trade with B and C. In the market on the right, if B and C reduce their prices, A has no incentive to reduce its price, since its customers do not trade with B and C.

Figure 2: The trade network can be used to find market inefficiencies.

trading partners and react to changes in prices. However, we restrict our attention to a fixed network that corresponds to a snapshot of the market during a particular interval of time.

The trade network of a market is likely to be sparse—most nodes will trade with a relatively small number of nodes. Individuals tend to trade with local businesses more than they do with businesses located far away. Firms tend to trade with other firms with whom they have prior relationships, and large contracts are not often renegotiated. So, firms will tend to be sparsely connected in the market. Many trading relationships—like subscription to cell phone service or rental agreements—are maintained with contracts that deter participants from switching trading partners. Even markets that may seem well connected at first, like the stock market, may not turn out to be so, on closer inspection. Traders in the stock market trade only with other participants that either hold the same stocks, or are interested in the same stocks. If two publicly traded firms A,B have businesses that operate in the same space, yet traders that own stocks of A are not well connected to traders that own stocks of B in the trade network, the market may have an inefficiency. This is the kind of inefficiency that our model finds.

It is worth noting that one network can be better connected than another, even though both have the same volume of trade. For example, Figure 2a shows two trade networks, each involving 12 participants. Every edge has the same rate of trade. In both cases, every participant is connected to 3 others, but the network on the left is better connected. There are only two edges crossing the gray divider on the right, so information that is known to the nodes on one side of the gray divider can take a long time to propagate to the other side. This illustrates the fact that measuring the connectivity of trade networks is not as straightforward as counting the total volume of trade, or estimating the number of participants involved in a lot of trade. The structure of the connections between the participants is important. The structure also reveals nodes that have undue control of prices because of their position. Figure 2a illustrates this using two networks where the nodes A,B,C are service providers that use contracts to restrict the ability of customers to switch providers. A,B and C have the same share of the trade in both examples, so market share alone cannot discern a difference between the two networks. However, in the market on the right, few customers trade with more than one provider. So, if B,C reduce their prices, A has very little incentive to follow suit. In the market on the left, customers do often trade with multiple providers, leading to a more efficient market. Observing the trade network can reveal such inefficiencies even though measuring market share does not reveal them.

A key difference between our approach and past work is that we do not attempt to describe the evolution of prices and do not attempt to estimate what prices ought to be. Instead, we measure the connectivity of the market by computing the effect of potential *shocks*. A shock is a change in the beliefs of market participants about the value of the products that they trade. For example, if many corn farmers experience a loss of their crop due to infestation, that event generates a shock in the market—some subset of market participants would then believe that corn is relatively more valuable. A *significant* shock is one that substantially changes the beliefs of many participants in the market. In an efficient market, a large volume of trade should be affected by any significant shock, leading to pressure on the participants to readjust their beliefs to a new norm. We give 3 definitions to quantify market connectivity:

**Tension and Energy** Intuitively, the *tension* of the network is the total normalizing force that market participants experience after any significant shock, and the *energy* of the network is the minimum amount of work that needs to be done to overcome market forces and bring the market to the state of any significant shock. Suppose two participants trade an item at rate $r$, and after the shock, their beliefs for the amount of the item that is worth a unit of currency are $x$ and $y$. Then the *force* generated by the shock is proportional to $r \cdot |x - y|$. The *energy* stored in the edge is equal to the amount of work that needs to be done to bring the market to this state: $(1/2) \cdot r \cdot (x - y)^2$. There is no force or energy if both nodes have the same beliefs. The force and energy are maximized when the difference in their beliefs is large and the rate of trade is high. The tension of the shock is the magnitude of the net force felt by all nodes. Up to normalization factors, the tension/energy of the network is the minimum over all significant shocks, of the tension/energy experienced by the nodes after the shock.

The tension and energy of a shock do have several qualitative differences. The tension is the total net force felt by all market participants, which corresponds to quantifying how quickly small changes in the beliefs of nodes helps to bring the nodes closer to having the same beliefs. The energy corresponds to quantifying the total potential energy stored in the edges when the nodes experience the shock. The energy is large when most nodes trade with partners that have conflicting beliefs from themselves. The tension is large when most nodes have beliefs that differ from the average of the beliefs of their trading partners.

Two large countries that do not trade correspond to a trade network with low tension/energy. Prices in one country can easily stray far from prices in the other country. These concepts can be used to distinguish the two examples shown in Figure 2a.

**Bottlenecks** A *bottleneck for tension/energy* is a small set of nodes that is responsible for a large fraction of the tension/energy of the network after some significant shock. Market participants that correspond to the bottleneck have an incentive to engage in anticompetitive practices, blocking the flow of information and interrupting the transmission of shocks.

A firm that has a monopoly on the production of an item that cannot be easily substituted with items available from competitors is a particular kind of bottleneck, but not the only one. Prices can be held at artificial levels by such a bottleneck. The concept of bottlenecks can be used to distinguish the two examples shown in Figure 2b.

A market with high *connectivity* is a market that has high energy, high tension, and no bottlenecks. The advantage of our theory over prior work is that it does not use an asset pricing model, so statements analogous to the efficient markets hypothesis are falsifiable in our theory. One can use our definitions to explicitly find and quantify inefficiencies if they exist. An inefficiency in our theory corresponds to a significant shock with low tension/energy, or a significant shock and a small set of nodes that form a bottleneck for the shock. Such an inefficiency identifies an obstacle for the flow of information that may be exploited for profit, reducing the inefficiency, or addressed by a change in policy or regulations.

In the restricted case where all trade in the market involves only one type of item, the formulas we use for tension and energy have long been used to model many physical systems, and are closely related to concepts studied in spectral graph theory [Spi, Chu96]. The same equations describe the behavior of electrical networks (where beliefs correspond to voltages, tension corresponds to the current flowing into the network, and energy corresponds to the power of the network), networks of springs, as well as the flow of heat in conducting materials. The fact that these equations have proven their worth in so many diverse settings is evidence that they capture something fundamental about the trade network. Viewed as a spring system, the rate of trade along an edge corresponds to the spring constant of the spring that it represents as in Hooke's Law. Each node's belief corresponds to a location on the number line. The placement of the nodes at these locations by the shock induces forces at the nodes, and the tension is the total magnitude of all these forces. The energy is the total potential energy stored in the system.

When multiple items are traded, the formulas we develop are natural multidimensional analogs of their single item counterparts—they loosely correspond to what happens when nodes are placed in a multidimensional space and the directions of the forces are twisted according to the type of item being traded. We associate each item with a vector that we call the *type vector* of the item. Similar items have nearly identical type vectors, and items that are different from each other have nearly orthogonal type vectors. We show how to use data from the trade network to generate the type vectors. Each edge applies a force to a node only in the direction of the type vector. These choices allow us to model how trade in one item can carry information about a shock in another item. To the best of our knowledge, the mathematics we have developed for this general setting is new.

Our intuition is that the trade network actually encodes all significant channels of information, even those that are external to the market. Let us make a few points to justify this belief. First, when we refer to the "market", we mean the *whole* market. Prices on two different stock exchanges can be correlated even if stocks in one exchange cannot be traded on the other, because participants that trade on these exchanges are connected in the market via their trade in other items. Second, even if there are significant channels for the transfer of information that are truly external to the market, traders that act on this information will create a channel of trade that parallels the flow of information in the external channel, leaving a trace of that flow visible in the trade. For example, if traders in one country use the price of iron ore in another to make decisions about whether to sell or buy iron ore in their local market, then discrepancies in the price of iron ore between the two countries will give traders incentives to trade iron ore or products related to iron ore between the countries. Conversely, if trade between the two countries is impossible, then information about the price of iron ore in the other country is not actually a very useful source of information for the local trader. So, the extent to which trade connects different parts of the market is a good proxy for the extent to which those parts are able to communicate information relevant to prices. If a large number of people make trading decisions using information they obtain from the same public source, then there will be correlations in their trading decisions, which will lead to many of them being connected to each other via short routes of trade in the market. Consequently, this source of information is reflected in the trade of the market, and will be accounted for by the mathematics of the theory. Finally, the theory we develop is intended to quantify the macroeconomic health of the market, and the mathematics is tuned to finding large scale inefficiencies. A channel of information is important to our theory only if it affects decisions that lead to a significant volume of trade—all such channels are likely to be reflected in the trade network.

After briefly discussing related work, in Section 2 and Section 3, we explain the intuitions behind the choice of the model and develop the theory. We show how several ideas in economics can be better understood using the concepts we develop in our work. In Section 5, we discuss some open questions, and possibilities for future work that we find worthwhile.

### 1.2 Comparison with Related Work

The efficient markets hypothesis is of fundamental importance, and there is a vast literature of prior work that has sought to understand it. This literature has proven to be illuminating and important, both theoretically and practically, and has had a major impact on the views and practices of professional investors. It is responsible for bringing data driven methods to bear on understanding finance. We refer the reader to the reviews of Fama [Fam70, Fam91] for details. Empirical economists have used different types of tests to try and understand if the market is efficient, which have been categorized in [Fam91] as follows. In *tests for return predictability*, historical data are used along with an asset pricing model to compute estimates for the correct returns on investments, and these are compared to actual returns realized by the market. In *event studies*, the speed at which prices adjust to significant new events is measured. In *tests for private information*, researchers attempt to understand whether or not market participants have private information that is not reflected in prices. To conclude that the market is *not* efficient using such tests, one must be sure that the information gathered by the test is at least as comprehensive as the information known to the market, and that the asset pricing model used is more trustworthy than the market itself. Perhaps not surprisingly, most of these tests have concluded that the financial market is efficient. That said, these tests do provide valuable insights into the inner workings of financial markets.

The theory we develop here is based on intuitions from this past work, especially the notions of event studies and tests for private information. However, the mathematics and data used in our theory are quite different from past work. We do not attempt to make any quantitative statements about the evolution of prices, nor do we attempt to understand what information the market should take into account. Accurately modeling such phenomena is extremely difficult, and past work suffers from the joint-hypothesis problem because of this difficulty. Instead, we evaluate markets as systems for communication. Unlike past work, we give explicit ways to quantify market inefficiencies and find them if they exist, so market efficiency as we define it can be unambiguously determined. If markets do not have high connectivity, this can be conclusively proven to be the case within our theory, so our theory does not suffer from the joint-hypothesis problem that plagues traditional concepts of efficiency. The primary motivation for our work is to strengthen the concept of market efficiency by giving it these features.

A different line of past work has tested a consequence of the efficient markets hypothesis: no trader should expect to make unusually large returns from trading in the financial market. Again, this is a vast area that we cannot hope to summarize here, so we refer the reader to the survey [Mal03]. This work is also extremely useful in practice, but the fact that traders cannot expect positive returns is not conclusive evidence of market efficiency. Traders may have difficulty generating positive returns even in an inefficient market due to high transaction costs, the difficulty of finding inefficiencies that they are capable of exploiting before they disappear, the lack of capital to compete with an entrenched system, or compliance with regulations.

Perhaps the biggest challenge to the efficient markets hypothesis has come from the field of behavioral economics [PS15]. Behavioral economics argues that the efficient markets hypothesis cannot hold because market participants are not rational, and are subject to human psychology. Our own model is built on the idea that market efficiency is limited by the bandwidth of information available to individual participants, so our work is inspired by similar considerations. However, unlike behavioral economics, our theory provides an alternative to the established theory of market efficiency that we believe has greater explanatory and predictive power.

There is a significant history of using networks to model social and economic phenomena. The book [Jac10] is a good reference. Several subfields here have developed very similar mathematics to the mathematics in our paper, but to the best of our knowledge, none are well suited to understanding market efficiency. One of the most related concepts from this past work is the *Bass* model of diffusion [Jac10, Chapter 7], which was invented to understand the spread of contagion in a social network. Contagion has been used to model the spread of financial crises [KRV03]. Here the emphasis is on understanding how a crisis spreads over financial relationships, rather than how useful information spreads, so the role of trade is quite different from in our model. Another body of work with similar mathematics arises in the study of learning networks [Jac10, Chapter 8]. In the DeGroot model [DeG74], every node of the network holds an opinion, and each node updates its opinion according to the opinions of its neighbors in the network. All of these past models are mathematically and conceptually related to our own model for the single item case, but do not appear to take into account the subtleties that arise from trade in multiple types of items. Indeed, the equations we use for the single item case have a much longer history, going back to the study of topics like electrical networks, networks of springs and the diffusion of heat. DeGroot also develops a notion of *social influence* that seems on the surface similar to our notion of bottlenecks, but is mathematically quite different. Social influence of a node in an *undirected* network is determined solely by the fraction of the total weight of edges that touch the node—the location of the node is not relevant. The concepts of *efficient networks*, and *graphical games* [Jac10, Chapter 6] are less related to the topic of this paper than they may seem from their names. They are concerned with understanding the space of strategies for playing games on networks.

## 2 The Trade Network for Single Item Markets

We use the notion of a trade network to understand the limits and power of markets to disseminate information via trade. The ideas we develop are simplest when there is only one item being produced and consumed in the market. We start by discussing this restricted case, because it is easier to understand. Later, we show how to model markets where an arbitrary number of items are traded,

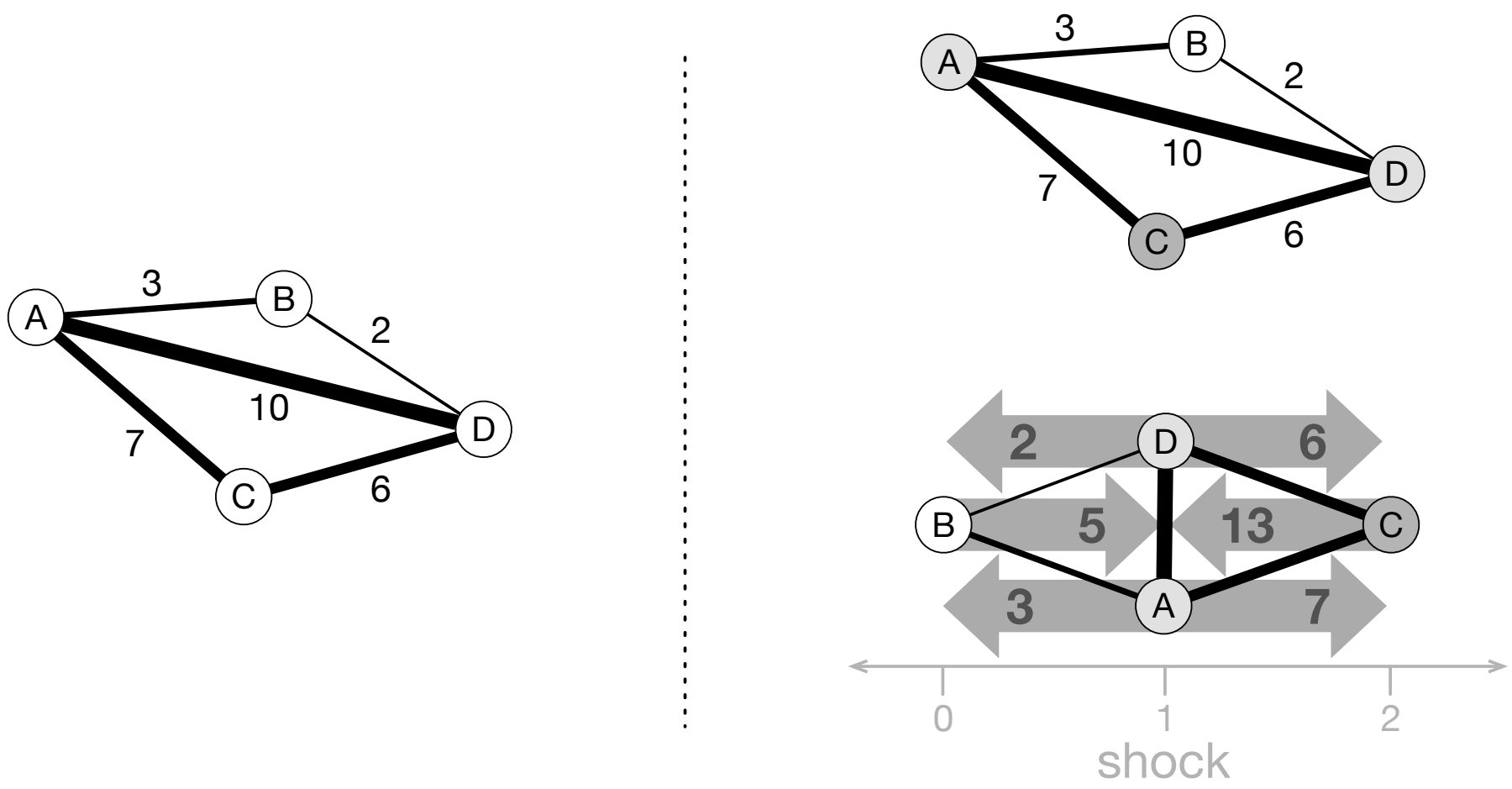

Figure 3: On the left: a single item trade network with 4 participants. The numbers on the edges denote the rate of trade. On the right: the network is shown with a mapping to the number line. The locations of the nodes correspond to a shock: a change to beliefs about the amount of the item that is worh a unit of currency. The arrows show the forces induced on each node by trade.

which is much more applicable to real markets.

Recall that we represent every market participant by a distinct node, and connect two nodes by an edge if the corresponding market participants directly trade with each other. We use $V$ to denote the set of nodes and $E$ to denote the set of edges. Every edge $e = \{u, v\}$ connects two nodes, $u$ and $v$ of the network. The edge is assigned a non-negative number called the *weight* of the edge, denoted $w(e)$, or $w(\{u, v\})$. The weight is the rate of the trade between the two participants. If $w(e) = 0$, that is equivalent to there being no trade along the edge $e$. To estimate the weight of each edge in real world markets, one can use past transactions to compute the monetary value of the trade between every pair of participants over a fixed time interval and then divide this number by the length of the interval.

Since single item markets correspond to spring systems, we start by giving some intuition[3] about the correspondence between the two. It is helpful to think of placing each node at the location on the number line that corresponds to its belief for the value of the single item being traded in the market. In equilibrium, every node has the same belief for the value of the item, and there is no tension in the network. A *shock* is a change in the beliefs for the amount of the item that is worth a unit of currency. It can be represented by a $n$ dimensional vector of numbers $\boldsymbol{x} \in \mathbb{R}^n$, where $\boldsymbol{x}_v$ denotes the node $v$'s belief about the change. The shock induces forces in the trade network that helps to reduce the shock. The force on a node $u$ along the edge $\{u, v\}$ is equal to $w(\{u, v\}) \cdot (\boldsymbol{x}_u - \boldsymbol{x}_v)$—its magnitude is proportional to the product of the weight and the difference between the new beliefs of $u, v$ about the value of the item. The larger the net forces felt by the nodes, the faster the shock dissipates in the market. The shock will be resolved quickly if many participants with differing beliefs about the value of the item trade with each other at a high rate.

[3]This is just one possible interpretation for the model. The model can be interpreted in a number of different ways, each with its own merits.

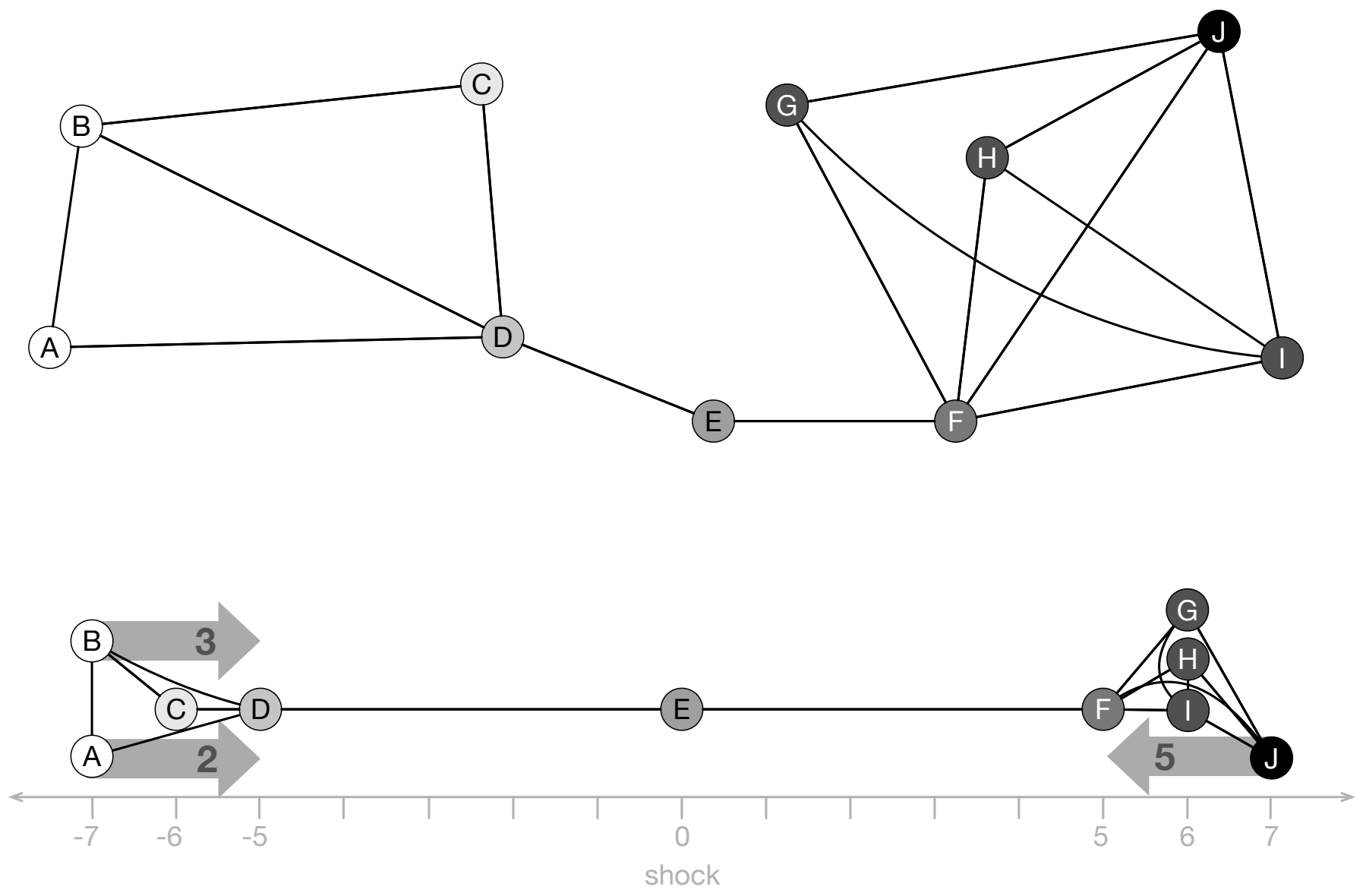

Figure 4: A mapping of a network to the number line. Every node is colored according to its location on the line, which is its belief about the amount of the item that a unit of currency is worth. The gray arrows show the forces experienced by nodes A, B and J. All other nodes experience no net force. The tension of the network for this shock is the sum of the magnitudes of the net force on each node, which is $\mathrm{T}(\boldsymbol{x}) = 3 + 2 + 5 = 10$. By Fact 2, the energy for this shock is $\mathrm{E}(\boldsymbol{x}) = 35$.

We are most concerned with the dissipation of shocks that are felt by a large part of the market.

## 2.1 The Laws of Ideal Springs

Next we review the basic laws of spring networks (the book [Bol98] is a good reference). Throughout the discussion we work with springs whose relaxed length is 0: namely the springs always pull nodes closer together and never push them apart. Hooke's law says the node $u$ experiences a force of

$$w(\{u, v\}) \cdot (\boldsymbol{x}_v - \boldsymbol{x}_u),$$

because of the edge $\{u, v\}$. The *energy* stored in the edge $\{u, v\}$ is equal to

$$\frac{w(\{u, v\}) \cdot (\boldsymbol{x}_u - \boldsymbol{x}_v)^2}{2}.$$

The energy is the amount of work that needs to be done to pull the two nodes of the spring apart. Given a shock $\boldsymbol{x}$, we define the tension induced by $\boldsymbol{x}$ to be the total magnitude of all the net forces felt by the nodes:

$$\mathrm{T}(\boldsymbol{x}) = \sum_{u \in V} \left| \sum_{v \in V} w(\{u, v\}) \cdot (\boldsymbol{x}_u - \boldsymbol{x}_v) \right|.$$

The total *energy* of $\boldsymbol{x}$ is the sum of the energies of each edge:

$$\mathrm{E}(\boldsymbol{x}) = (1/2) \cdot \sum_{\{u,v\} \in E} w(\{u, v\}) \cdot (\boldsymbol{x}_v - \boldsymbol{x}_u)^2.$$

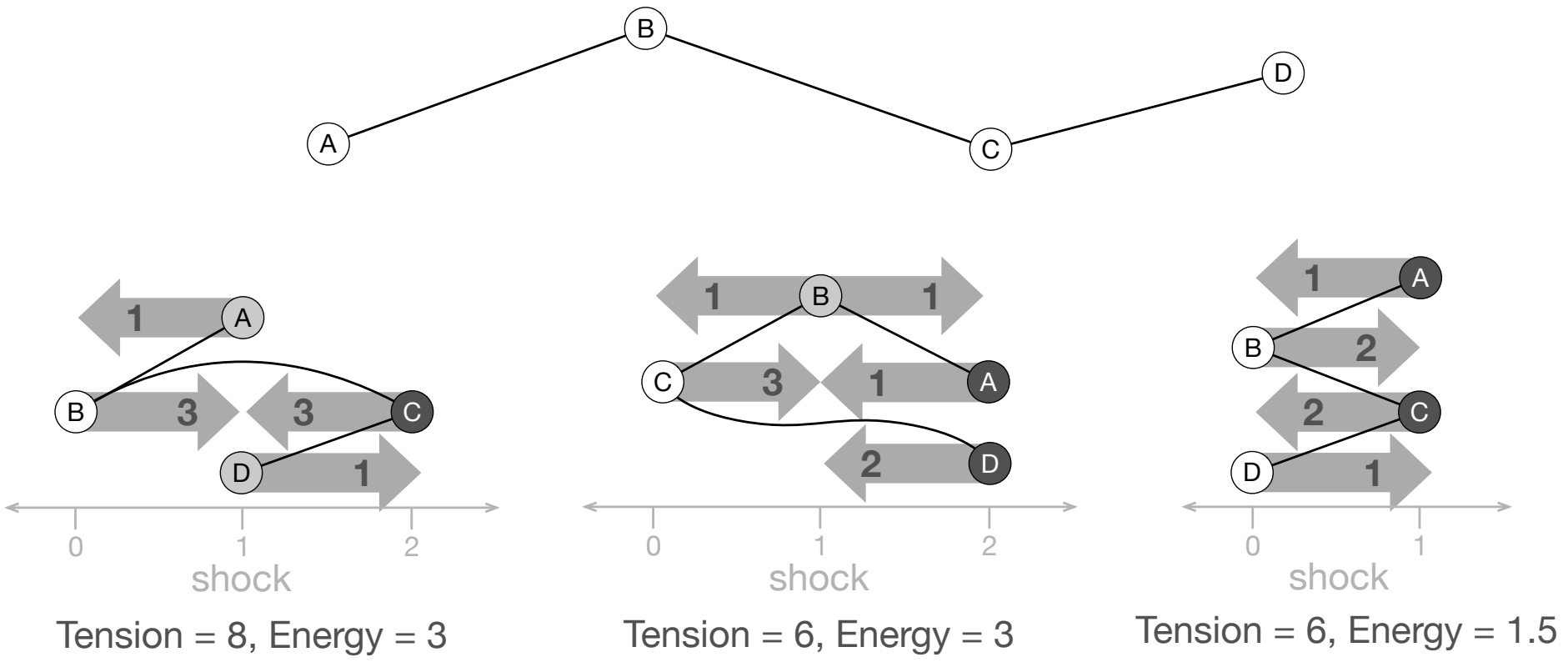

(a) Tension and energy are incomparable.

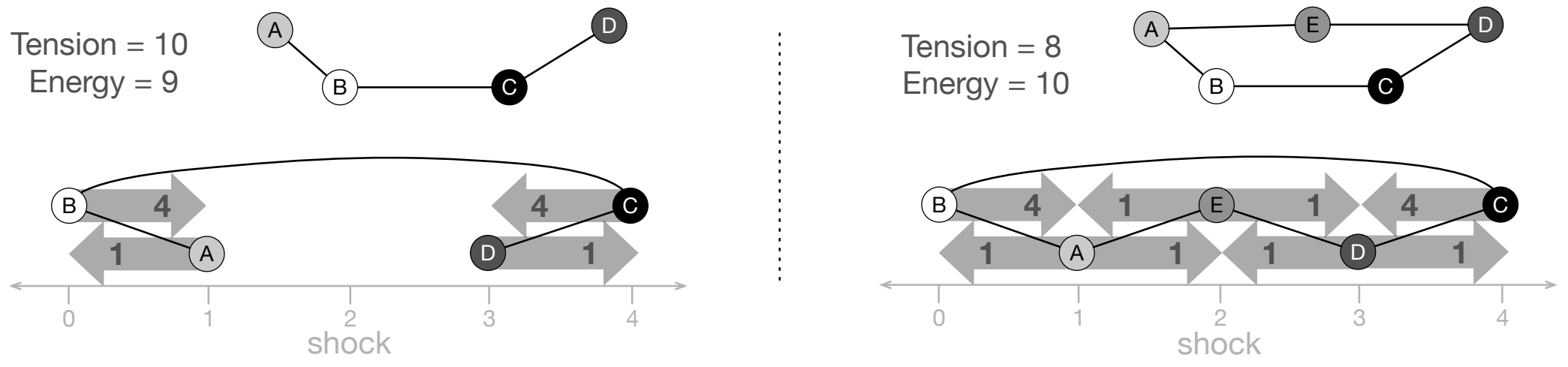

(b) The tension may decrease by trade with new nodes, but the energy does not.

Figure 5: Examples illustrating the difference between the tension and energy of a shock.

One can check that the gradient[4] of the energy is the vector of forces that need to be applied to the nodes to hold them in place:

$$(\nabla \mathrm{E}(\boldsymbol{x}))_u = \sum_{\{u,v\} \in E} w(\{u,v\}) \cdot (\boldsymbol{x}_u - \boldsymbol{x}_v), \tag{1}$$

so $-\nabla \mathrm{E}(\boldsymbol{x})$ is the vector of net forces experienced by the nodes, and the tension is the $L_1$ norm of the gradient vector:

$$\mathrm{T}(\boldsymbol{x}) = |\nabla \mathrm{E}(\boldsymbol{x})|_1 = \sum_{u \in V} |(\nabla \mathrm{E}(x))_u|.$$

The energy of the network is equal to the amount of work that needs to be done to bring the network to its current position from a 0 energy state. The following fact is useful to keep in mind to understand the examples we discuss next:

**Fact 2.** *If some subset of the nodes are held at location $a$ and some other subset are held at $b$, and the rest are left to settle at equilibrium, the energy is always equal to $|(a-b)t/4|$, where $t$ is the tension.*

[4]The gradient is the vector of partial derivatives.

The tension and energy have qualitative differences that are explored in Figure 5. In Figure 5a, two of the shocks have the same energy, and two of them have the same tension. The first shock gives the highest tension because it is the easiest for the participants to reduce the energy locally. In the middle example, node B cannot help to reduce the energy. This is what leads to the second example having lower tension than the first. In the final example, the energy is much lower because the participants are much closer to having the same beliefs than in the other two examples. Loosely speaking, the tension corresponds to the ability of most participants to reduce the distance to equilibrium, while the energy corresponds to the total work that needs to be done to return the market to equilibrium. In Figure 5b, we see that the presence of a new node can actually decrease the tension of the network, though it can never decrease the energy. For simplicity of explanation, these subtleties are ignored in the examples we discuss next. The shocks in all of these examples have been chosen so that they do not distinguish between tension and energy.

## 2.2 Motivating Examples

The model we define in this paper can be used to explain diverse phenomena that pertain to market efficiency. Here we consider several examples of well known ideas in economics that can be explained using our model. We stress here that a proper treatment of actual markets should take into account that the items traded in the market are different from each other. We discuss the theory that handles this more realistic setting in Section 3. To compute the connectivity of the market, we find significant shocks that have low energy, low tension or reveal the presence of bottlenecks.

Suppose the market under consideration consists of $n$ farmers that produce and consume corn. Every farmer buys and sells corn by trading in a common market place, and every farmer trades with every other farmer at a rate of 1. Then every pair of nodes in the trade network is connected by an edge of weight 1. If a single farmer produces less corn than expected, the effects of this shock are quickly felt by the whole market. The farmer will seek more corn from others, which increases the price of corn and incentivizes the others to produce and sell more corn to compensate. If a large fraction of the farmers experience a drop in production, the effect on the network is even more pronounced. Viewed as spring network, where every edge has weight 1, we see that the tension between any two disjoint sets $S, T$ that have different values in the shock is quite large. For example, suppose that $S, T$ are each of size at least $n/3$, every node in $S$ has a value of $+1$, and every node of $T$ has a value of $-1$. Then there are at least $n^2/9$ edges going directly from $S$ to $T$, so the tension of any such shock is at least $4n^2/9$, and the energy is at least $2n^2/9$. So every shock gives high energy and tension, and the market has high connectivity.

It is a commonly held view that free trade increases market efficiency, a view that can be quantified with our model. Suppose that the market consists of farmers living in two adjacent towns, each with $n/2$ farmers that do not trade with each other. If the farmers in one of the towns experience an oversupply of corn, the price of corn, and so the economic activity, remains unaffected in the other town. Viewing the trade network as a network of springs, as in Figure 6a, we see that the tension and energy are minimized if every node in one town is placed at $-1$, and every node in the other is placed at 1. For this shock, we have $\mathrm{T}(\boldsymbol{x}) = \mathrm{E}(\boldsymbol{x}) = 0$. This matches the intuition that there is no way for information to flow between the two towns via trade. It will not be long before traders emerge between the two towns. Consider now the trade network depicted in Figure 6b, where there are three traders trading corn at a rate of 1 along a road connecting the two towns. This market is more efficient than the disconnected market that we had before, since trade

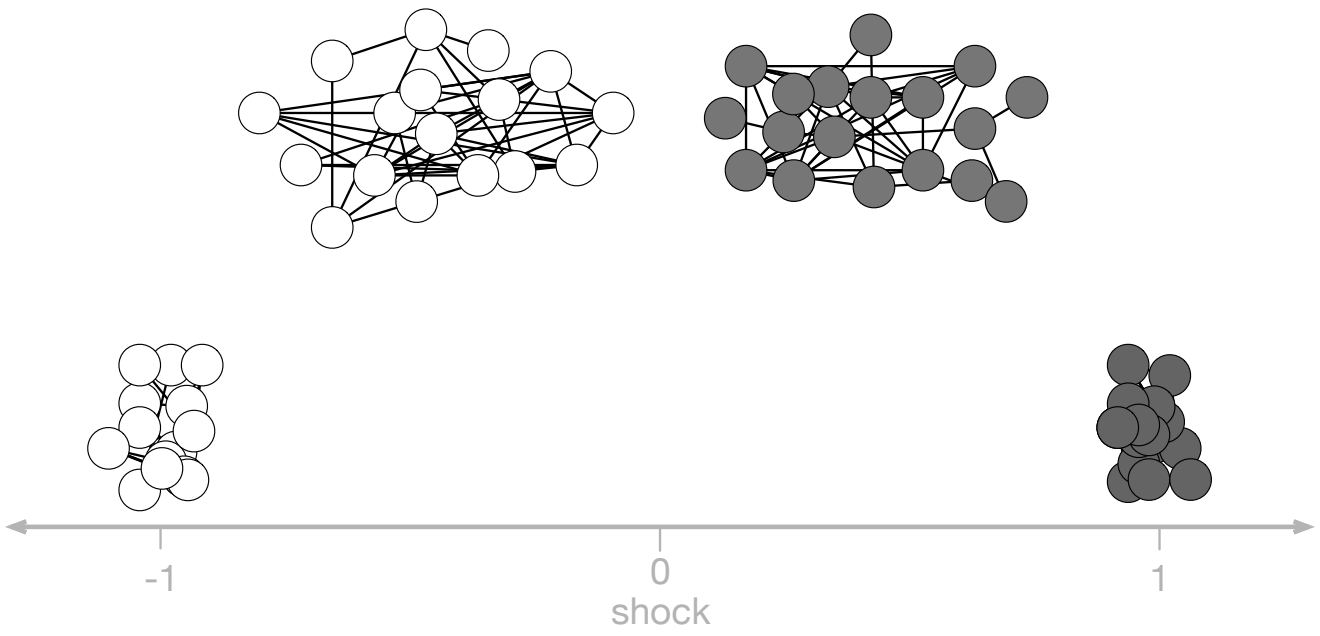

(a) $\mathrm{T}(x) = \mathrm{E}(\boldsymbol{x}) = 0$ when all the white nodes in one part are placed at $-1$, and all the gray nodes from the other part are placed at 1. The shock identifies the disconnection in the network.

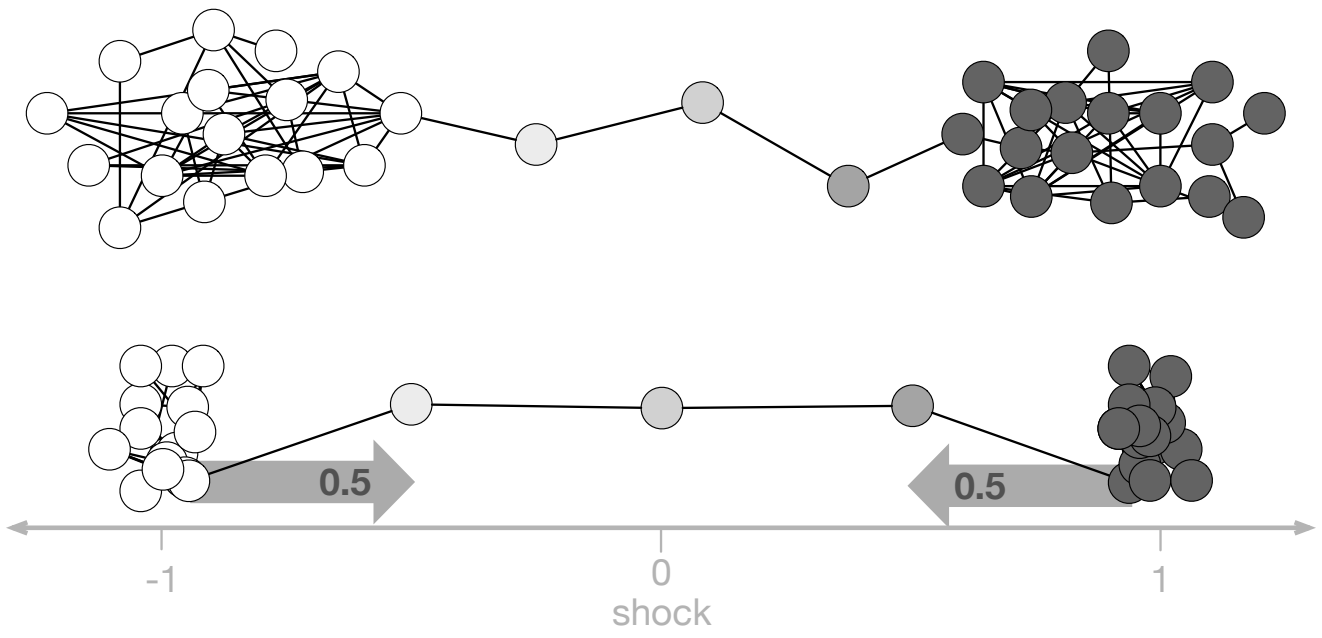

(b) Only the nodes at $-1$ and 1 experience a net force. The tension and energy of the shock has gone up to $\mathrm{T}(\boldsymbol{x}) = 1, \mathrm{E}(\boldsymbol{x}) = 0.5$ from Figure 6a.

Figure 6: All edges have unit weight in this sequence of examples showing that tension and energy increase with connectivity.

does communicate information between the two towns. In line with our intuition, if every node in one town is placed at $-1$, and every node in the other is placed at 1, and the rest of the nodes are allowed to relax at equilibrium, a force of 0.5 pulls the nodes on the left towards the nodes on the right. The tension is 1 and energy is 0.5. Now suppose that there was just 1 trader between the two towns as in Figure 7a. Then the market would be better at communicating information from one town to the other, since only one trader is involved in communicating the shock from one town to the other. This is reflected in the fact that the tension of the shock has increased to 2. The market is most efficient when there are multiple independent traders trading between the two towns, as in Figure 7b. The tension of the shock has increased to 6.

Trade networks can be used to explain the success of many modern innovations in finance. Consider the role of *exchange traded funds* in the stock market. These are funds that hold assets like stocks or bonds, and allow investors to trade financial instruments that correspond to diversified portfolios. In recent years, the volume of trade in exchange traded funds has risen dramatically. The total annual trade in these funds exceeds the U.S. GDP, and the funds have a turnover rate that is more than 4 times larger than average stocks [Bal15]. Our model suggests that this high level of trade contributes tremendously to increasing market connectivity. It is quite hard for a

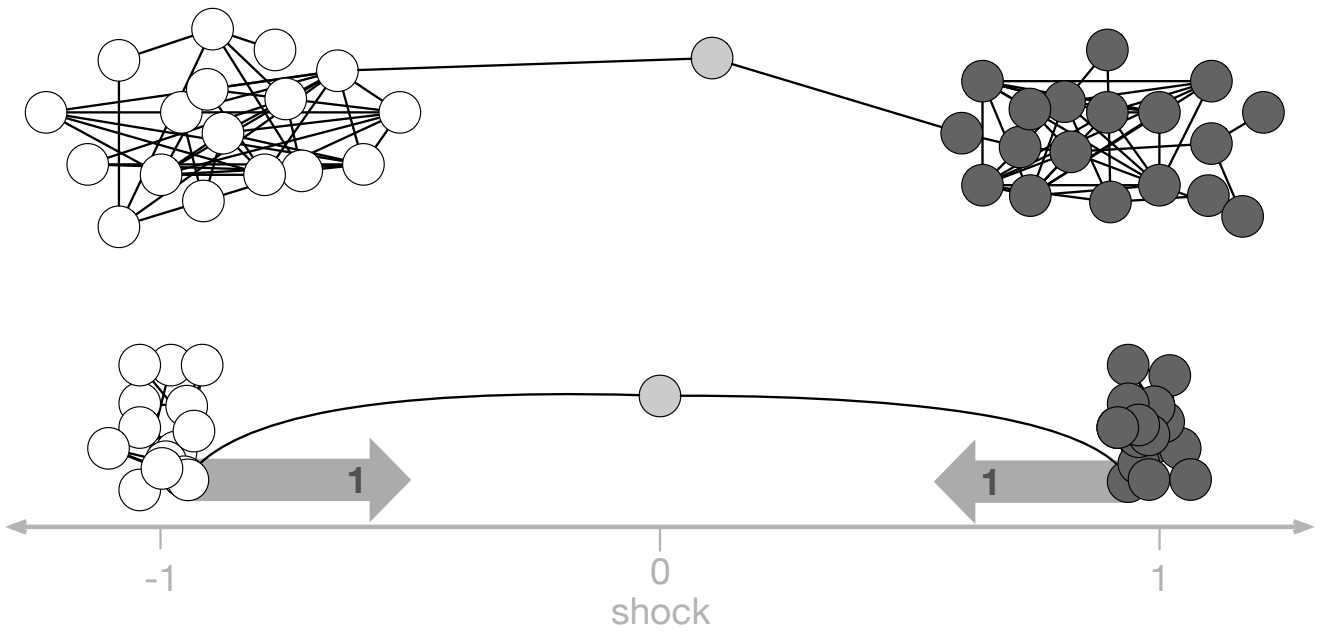

(a) A short trade route generates more tension than the trade route in Figure 6b. $\mathrm{T}(x) = 2, \mathrm{E}(\boldsymbol{x}) = 1$. The node at 0 experiences no net force.

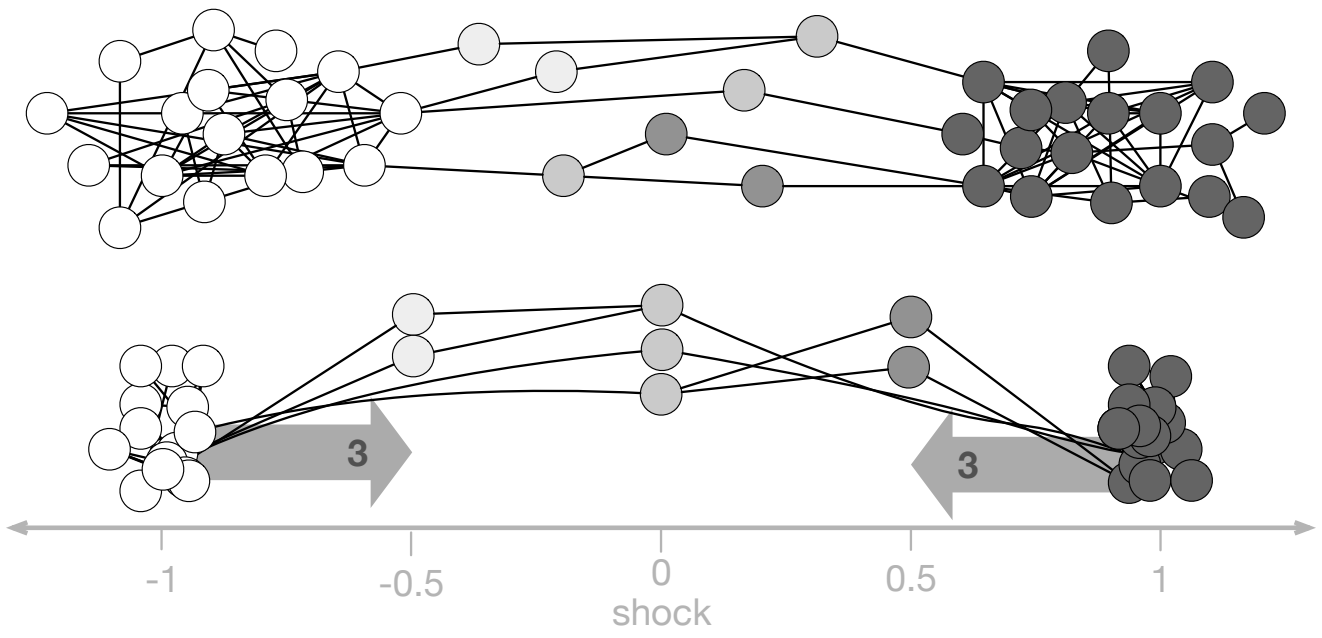

(b) Many parallel trade routes generate more tension than in the network of Figure 7a. $\mathrm{T}(x) = 6, \mathrm{E}(\boldsymbol{x}) = 3$. Only the nodes at $-1$ and 1 experience a net force.

Figure 7: All edges have unit weight in this sequence of examples showing that tension and energy increase with connectivity.

typical investor to monitor and trade a truly diversified portfolio of stocks, so if these (or similar) funds were not available, investors would be trading only with others that are interested in the same stocks. Since the set of participants interested in trading a highly diversified exchange traded fund is much larger, trade in these funds makes the trade network much more connected than before. Investors that were formerly focused on disjoint portfolios that are nevertheless associated with the same exchange trade fund will now begin to trade with each other. This makes the trade network much more likely to have high energy and tension for a given shock[5]. Exchange traded funds offer a channel of communication that reaches a much broader set of participants than trade in the constituent stocks can provide.

A popular view in economics is that the availability of liquid assets is vital to the functioning of a healthy economy. For example, Chordia, Roll and Subrahmanyam write "Liquidity facilitates efficiency, in the sense that the market's capacity to accommodate order flow is larger during periods when the market is more liquid." [CRS08] This view is supported by our model as well. A market with high liquidity is one in which nodes are more likely to be connected to many trading partners via trade. Such a network is much more likely to have high energy and tension than a sparsely connected market.

[5]If two publicly traded companies operate in the same space, but most shareholders of one company do not hold shares of the other, our model suggests that the introduction of an exchange traded fund giving exposure to both companies can improve market connectivity.

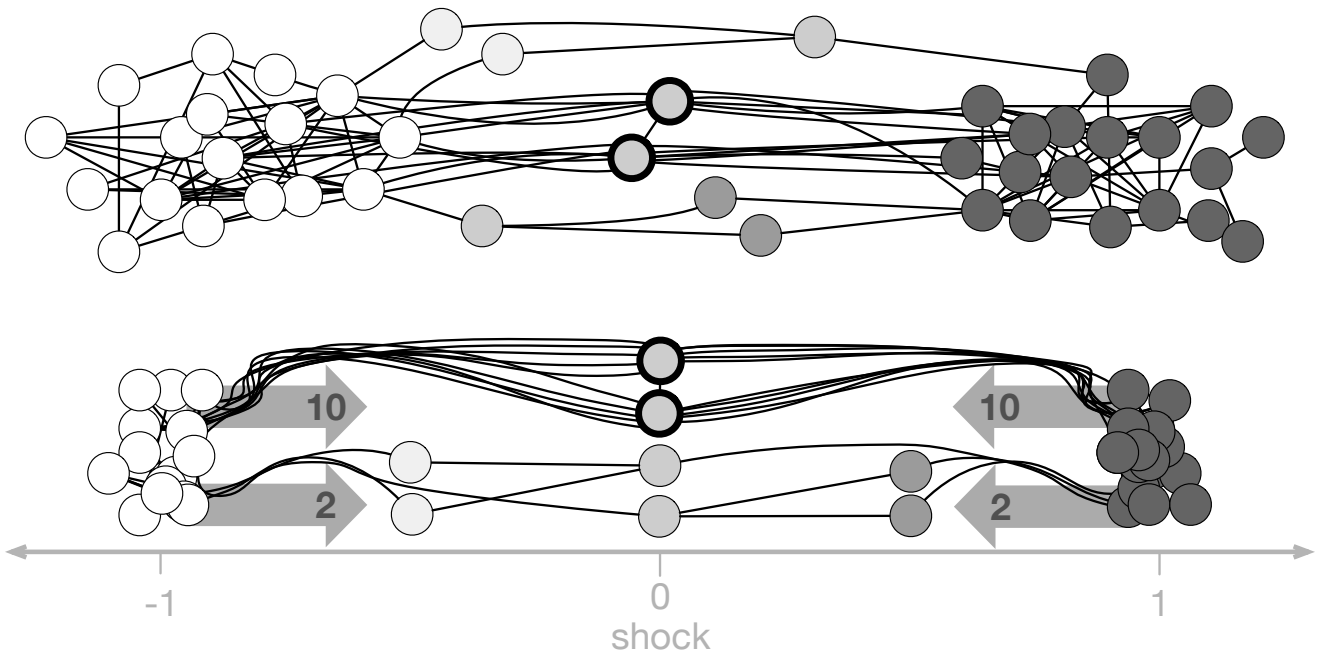

Figure 8: Two nodes that are responsible for most of the tension/energy form a bottleneck.

A second kind of market inefficiency that we study corresponds to monopolies in the market. Suppose a small set of traders comes to control a very large fraction of the trade between the two towns, as in Figure 8. We have $\mathrm{T}(\boldsymbol{x}) = 24, \mathrm{E}(\boldsymbol{x}) = 12$, but the tension and would drop dramatically to $\mathrm{T}(\boldsymbol{x}) = 4, \mathrm{E}(\boldsymbol{x}) = 2$ if the highlighted nodes are removed. This is what we call a *bottleneck* in the trade network. It is a small set of nodes that contributes most of the tension/energy.

In some settings, the set of nodes involved in the bottleneck have an incentive to commit fraud, or engage in monopolistic practices, because they have a large market share. Even if these nodes behave in good faith, information flow in the market relies too heavily on the bottleneck, which can lead to the market being inefficient if these nodes make errors in their adjustments to prices.

An extreme example of a bottleneck can be observed in the trade network of a centralized economy, shown in Figure 9. This network is well connected. In fact, from the perspective of the tension and energy, it is nearly as goodf as the fully-connected network, because there is a short path of length 2 connecting every two nodes in the network. However, every individual trades only with the central authority, depicted as the node in the middle. Even though the tension and energy of the network are quite high, the network clearly has a bottleneck. The burden for recognizing and transmitting shocks falls entirely onto a single entity, making it unlikely for information to be reliably communicated.

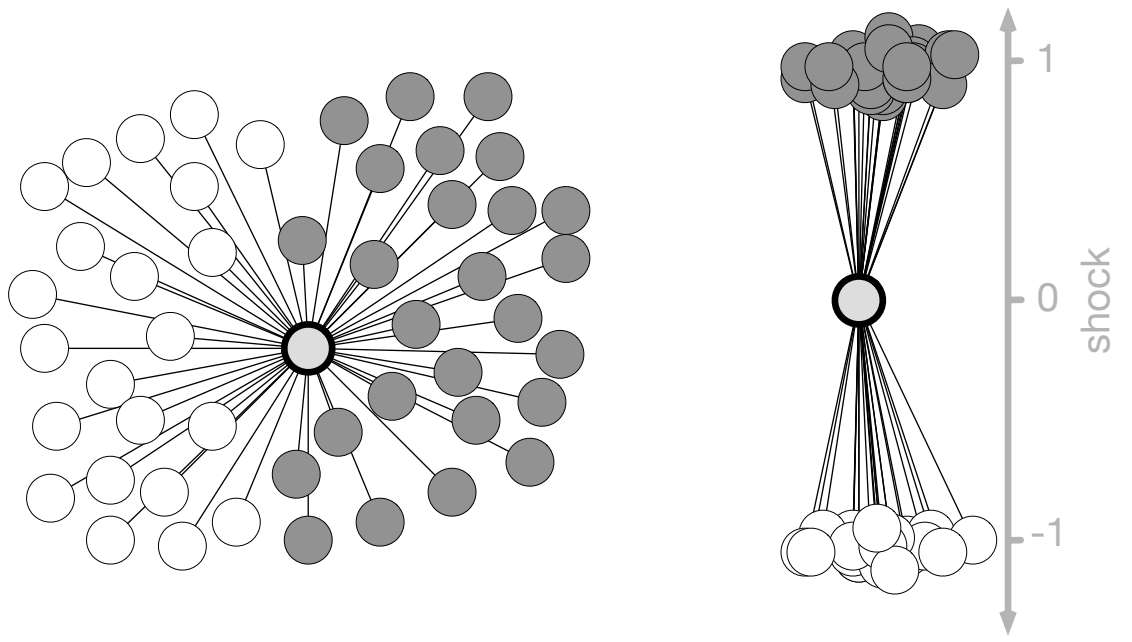

Figure 9: In a centralized economy, every node trades only with the central authority, which is a bottleneck.

It is widely recognized that the financial crisis of 2007-2009 began because of errors in accurately pricing mortgage backed securities (MBS). A crash in the value of these securities initiated a global recession. Data gathered about the organization of the market prior to the crisis suggest that a bottleneck was to blame. As Fligstein and Goldstein report:

> Another commonly voiced myth about the MBS market is that it was highly dispersed, with too many players to control any facet of the market. On the contrary, we show that over time all of the main markets connected to MBS, the originators, the packagers, the wholesalers, the servicers, and the rating companies became not only larger, but more concentrated. By the end, in every facet of the industry 5 firms controlled at least 40% of the market (and in some cases closer to 90%). Separate market niches also increasingly condensed around the same dominant firms. As a result, the mortgage field was not an anonymous market scattered across the country, but instead consisted of a few large firms. This concentration meant that firms very much collaborated and competed in these various markets. Firms would join together in MBS packages and assume different roles with each other. This meant that they had a great deal of knowledge of the market and what the others' moves were. [FG10, p. 33]

One of the major goals of our work is to develop tools that can be used to detect structural problems like bottlenecks before a crisis occurs. Economists and regulators could use the theory to identify bottlenecks in the trade network and investigate methods to tackle them before they lead to a crisis.

Market connectivity can also be used to assess financial solutions that mitigate the effects of bottlenecks. For instance, consider the role of *futures* in the commodities market. Futures contracts guarantee delivery of a commodity at a date in the future. Perhaps surprisingly, it is often the case that only a small proportion of contracts that are traded are settled by actual delivery. Our model suggests an explanation for this counterintuitive fact: futures contracts serve as a mechanism to break bottlenecks in the trade network, and they are most effective when the trade in the futures exceeds trade in the actual item. We illustrate this using the following anecdote by Hieronymus [Hei77] about the origins of futures contracts in corn:

> A country merchant rides into town in January and proceeds to the place of business of the terminal merchant to whom he regularly sells. On offering 20,000 bushels for June delivery at the current price he hears, "I would like to bid that much but with the large stocks in Chicago and a large crop coming in I can only pay 15 cents below today's price." Our friend mentions the high price that he has already paid farmers, comments on the ancestry of terminal grain merchants in general, and takes himself to the nearest saloon to find solace. There he comments to all and sundry regarding the greed and cowardice of grain merchants in general and one in particular. On hearing this one stalwart soul says, "I know nothing of corn, being a builder of houses myself, but it occurs to me that the price of corn will be quite as high in June as it is now." Being true to his occupation, just as we know country merchants today, our man asks, "Is that a firm offer?" "I shouldn't want to go quite that far but I will bid five cents below today's price. That will take you off of the hook and out of your cups and leave room for a bit of a profit for me." "Done," replies the country merchant and they sign a contract. Some weeks later the builder of houses, who has now become interested in these matters, notes that the price of corn for June delivery is five cents above the price that he has paid. He figures that $1000 in his pocket is better than 20,000 bushels of corn in his lap in June so he peddles his contract to the nearest terminal merchant and wonders why he had not discovered this easy road to riches sooner. [Hei77, p. 74]

The transformation to the trade network described in this anecdote is shown in Figure 10a. The country merchants are represented by the nodes A and B, and the terminal merchants by the nodes

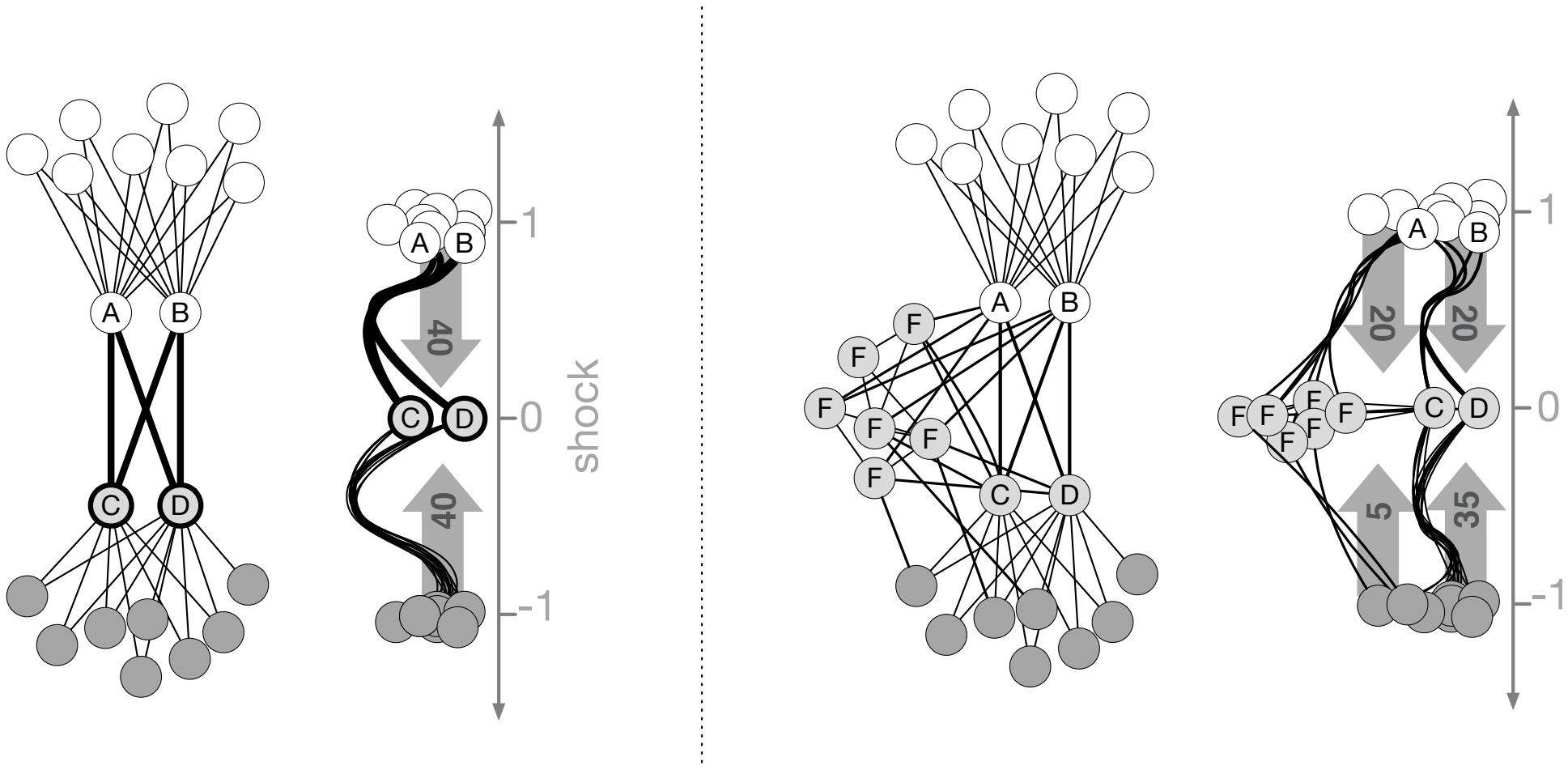

(a) A futures market can help to break a bottleneck. On the left, the merchants A and B can only trade with merchants C and D. C and D are a bottleneck for the depicted shock. The introduction of futures traders labeled F substantially changes the network to the figure on the right. A,B can now sell their contracts to the futures traders, which reduces the role of C,D in generating the tension in the network. C and D must now compete with the traders F for the contracts and cannot fix prices.

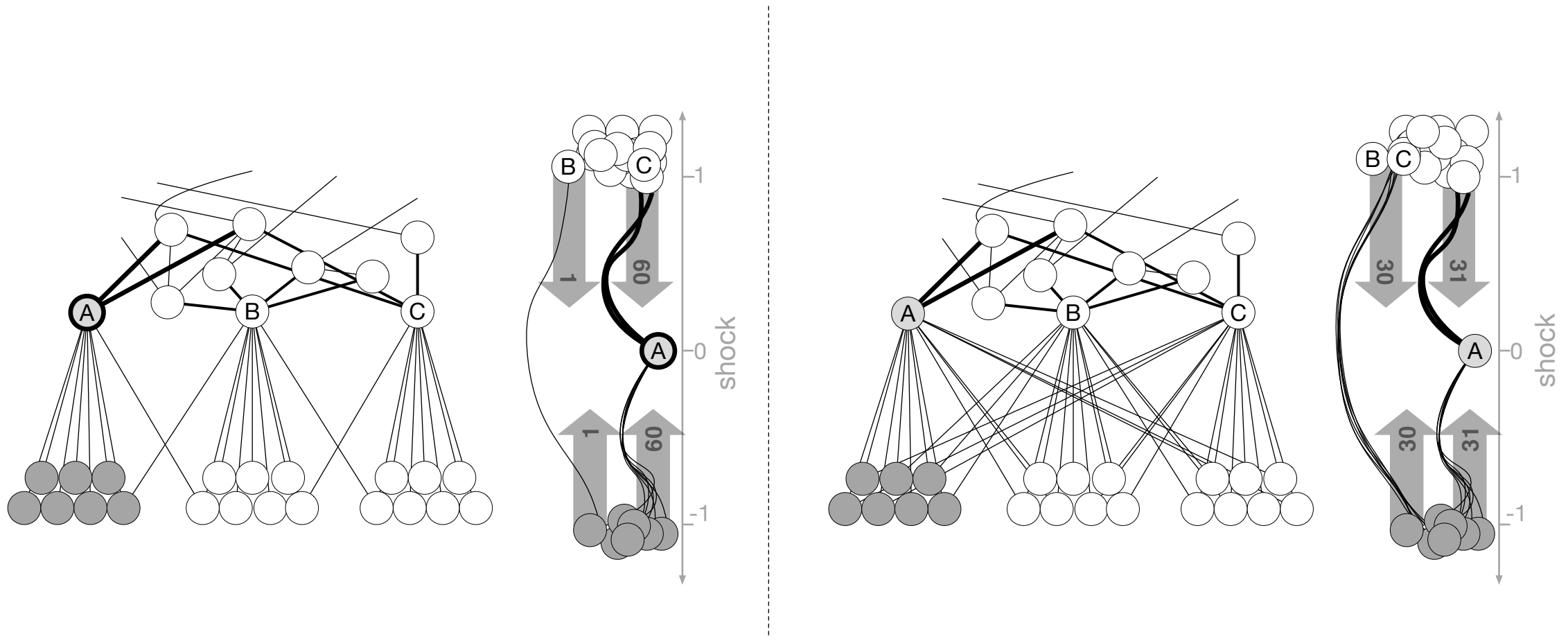

(b) High switching costs can give rise to trade networks with bottlenecks. In the network on the left, A,B,C provide service to a set of customers that correspond to the nodes on the bottom. Switching costs are high, so most customers trade with a single provider. A is a bottleneck for the depicted shock. If switching costs are low, we might see the network shown on the right, where A is no longer a strong bottleneck.

Figure 10: Bottlenecks explain the impact of high switching costs and the benefits of a futures market.

C and D. If futures contracts were not available, the network would look like the one on the left, where C and D clearly form a bottleneck for the depicted shock. They are able to charge unfair prices from A and B because A and B cannot sell their corn to anyone else. The introduction of the futures market dramatically alters the trade network, as shown on the right. There are two reasons why C and D become less of a bottleneck according to our definitions. The first is more obvious: the tension and energy that they are responsible for has been reduced. For the given shock, C and D are responsible for only a small part of the tension, since a sizable fraction of the trade now involves the futures traders instead of C and D. Consequently their ability to extort higher prices has been greatly diminished. A second reason is that the significance of the depicted price shock has been reduced, because a larger fraction of the traders have the same beliefs in the depicted shock. A shock is now significant only if the futures traders have significant disagreement about the values in the shock. C and D cannot manipulate prices very easily because the high volume of trade in futures contracts will reveal accurate prices for corn.

Next we consider the effects of switching costs on market efficiency. Markets where participants are locked-in to trade with particular nodes will tend to be less efficient than markets where participants can freely switch trading partners. As Farrell and Kemperer observe,

> Switching costs [...] bind customers to vendors if products are incompatible, locking customers or even markets in to early choices. Lock-in hinders customers from changing suppliers in response to (predictable or unpredictable) changes in efficiency, and gives vendors lucrative ex post market power [...] [FK07, p. 1970]

The effects of two types of switching costs can be seen in Figure 10b. The figure shows three service providers, A, B, C that serve a group of customers, represented as the nodes on the bottom. If switching costs are prohibitively high, one might observe the network shown on the left. The typical customer trades with only one of the service providers. If a shock occurs where the customers of A have a different belief from the rest of the market, A becomes a bottleneck for the shock. The network on the right shows what the network might look like if switching costs are lower. Now most customers trade with multiple providers, and we see that A is less of a bottleneck for the same shock, because customers also experience tension and energy because of the edges that connect them to B and C. A can no longer fix prices because the competition with B and C is much more of a threat. Note that this example shows that trade networks are more effective at identifying monopolies than counts of market share. In this last example, A has only a 1/3'rd share of the market. However, its ability to lock-in customers gives it a considerable advantage in the market.

Another interesting example of the explanatory power of trade networks comes from understanding *vertical integration*. Vertical integration refers to an arrangement where a firm owns several parts of the supply chain of a product. Does market efficiency improve with vertical integration? On the one hand, vertical integration does increase the energy and tension of most shocks; on the other hand, it has been observed that vertical integration is associated with increased rates of fraud. Fligstein and Roehrkasse [FR16] make the case that vertical integration was to blame for the elevated levels of fraud that contributed to the financial crisis of 2007-2009. They write,

> [...] [T]he motivation toward fraud increases with vertical integration because the diffusion of fraud is a precondition of continuing normal operations. [...] [T]he transaction points that vertical integration eliminates are not only sites of potential fraud. They also represent key opportunities for quality control and due diligence. [FR16, p. 624]

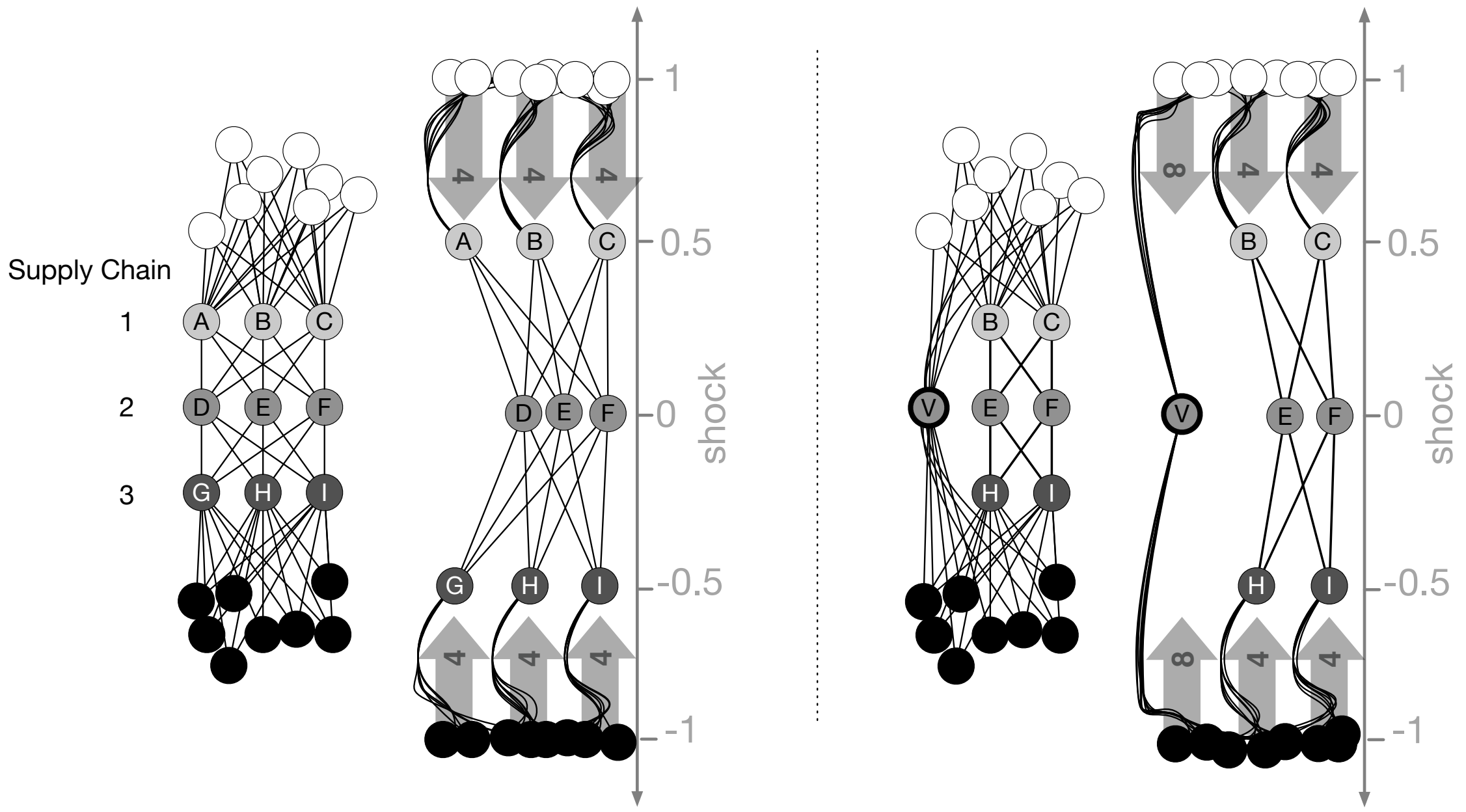

Figure 11: Vertical integration can lead to bottlenecks. The network on the left shows a supply chain involving the firms A, B, C, D, E, F, G, H, I. Each firm trades at an equal rate with the adjacent levels of the supply chain. The network on the right shows the same shock applied after firms A, D, G have been combined into the firm V. The tension and energy have increased, but V now contributes 1/2 of the tension and energy, even though it is involved in only 1/3'rd of the trade.

The effect of vertical integration is illustrated in Figure 11. The network on the left shows a supply chain connecting two parts of the network. Each step of the supply chain involves 3 firms with equal market share. If there is a shock between the two ends of the chain, each of the firms on the boundary of the supply chain contributes equally to the tension and energy. The network on the right shows what happens if three of the firms integrate vertically. The new firm has not increased its market share or the rate of trade that it is involved with, but its share of the tension and energy have increased dramatically. Even though it is involved in only 1/3'rd of the trade, it now contributes 1/2 of the tension and energy. This is because the edges associated with V have been stretched to be twice as long as the tension generating edges that V replaced.

# 3 The Trade Network for Multiple Items

The real power of markets emerges when one considers trade networks involving many different kinds of items. It is a mistake to view such a market as a collection of distinct markets trading each item, because shocks in one item can be dissipated via trade in another item. This makes multi-item networks more efficient than any of the subnetworks obtained by restricting attention to individual items.

Consider the market shown in Figure 12, which is a market of farmers that grow both corn and wheat. Viewing the market as a network for trading corn and a separate network for trading wheat suggests that the corn network has 0 tension and energy for a significant shock, since no corn is traded between the left and the right. But this is misleading. If the farmers on the left experience a sharp drop in the production of corn, the price of corn will rise. This will increase the demand for wheat on the left, making the price of wheat rise on the right, and eventually leading to a rise in the price of corn on the right—the trade in wheat transmits shocks in the corn market. However, if the trade between the left and right is in a commodity, say iron ore, that is not closely linked to corn, then the market is much less effective at communicating shocks in the corn network. These considerations motivate the choices we make next.

We associate every item traded in the market with a unit vector $\boldsymbol{\tau} \in \mathbb{R}^d$, for some number $d$. We refer to $\boldsymbol{\tau}$ as the *type* vector of the item. The type vectors are chosen in such a way that the inner product $\langle \boldsymbol{\tau}, \boldsymbol{\eta} \rangle$ captures the correlation between the items represented by the vectors $\boldsymbol{\tau}, \boldsymbol{\eta}$. So two items that are completely different from each other are represented by nearly orthogonal vectors—$\langle \boldsymbol{\tau}, \boldsymbol{\eta} \rangle \approx 0$—and two items that are very similar in terms of their role in the market are represented by vectors that are almost the same—$\langle \boldsymbol{\tau}, \boldsymbol{\eta} \rangle \approx 1$. Every edge $e$ of the trade network corresponds to an item, and we let $\boldsymbol{\tau}(e)$ denote the corresponding type vector[6]. To allow for the fact that two participants may trade multiple types of items, we allow the set of edges $E$ to contain multiple edges between two vertices $u, v$. The mathematics we develop has the feature that the values of the inner products between all pairs of items is the *only* relevant feature of the type vectors to the model. For example, rotating all the type vectors by the same degree will preserve all the quantities we define, because doing so does not change any of the inner products.

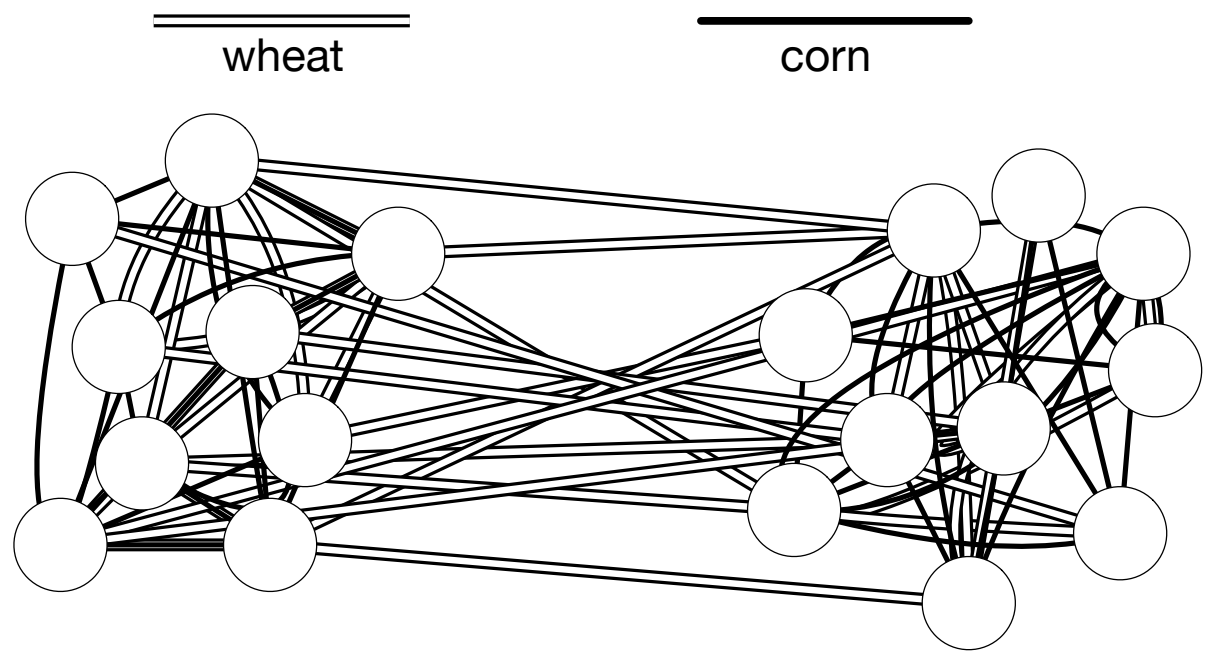

Figure 12: Trade in wheat can carry information about corn, so one should not view the trade network as two different trade networks for each item.

One can treat the vectors that represent items in the market as parameters of the model, or use the trade network itself to generate them. If the market has $n$ participants, and $E'$ denotes the subset of the edges that trade item $z$, let $\boldsymbol{\tau}'$ be the $d = n$ dimensional vector such that $\boldsymbol{\tau}'_v = \sum_{v \in e \in E'} w(e)$ is equal to the total amount of trade of item $z$ that node $v$ engages in. Then represent item $z$ using the unit vector $\boldsymbol{\tau} = \frac{\boldsymbol{\tau}'}{\|\boldsymbol{\tau}'\|}$. Two items that are largely traded by disjoint parts of the market will correspond to nearly orthogonal vectors, and two items that are often traded by the same nodes will correspond to vectors that have a large inner product. Two items that are substitutable will likely be sold and traded by the same nodes in the market, and two items that are unrelated to each other will likely be traded by different nodes. This choice corresponds somewhat to our intuitions, but may not always be ideal. One may need to come up with these vectors using a combination of data driven methods and information about the underlying items.

Note that if $I$ types of items are traded in the market, it is always possible to set $d \leq I$, since we can always project the type vectors down to an $I$ dimensional space while preserving the inner

[6]These type vectors can be used to encode the geography of the market even in single item markets. If we want to take into account the fact that items that are nearby are likely to be more correlated, we can associate every edge with a unit vector that is proportional to the vector encoding the location of the trade.

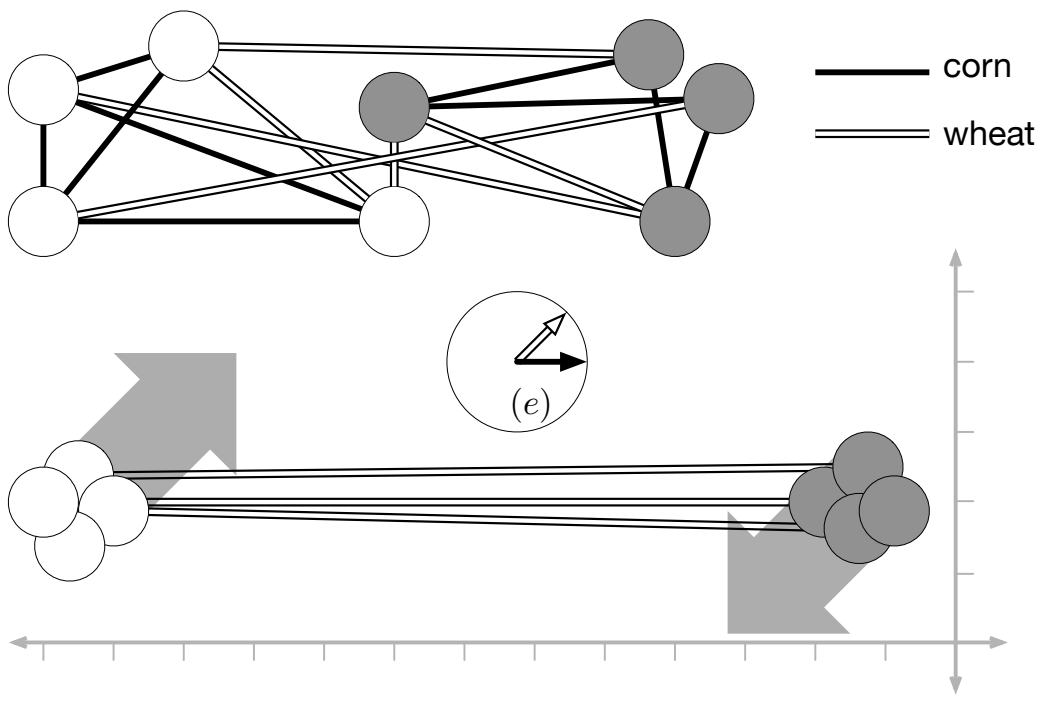

(a) If the white nodes are placed at the point $(-1, 0)$, and gray nodes are placed at the point $(1, 0)$, the market experiences some tension along the direction corresponding to wheat, because the type vector for wheat is not orthogonal to the shock.

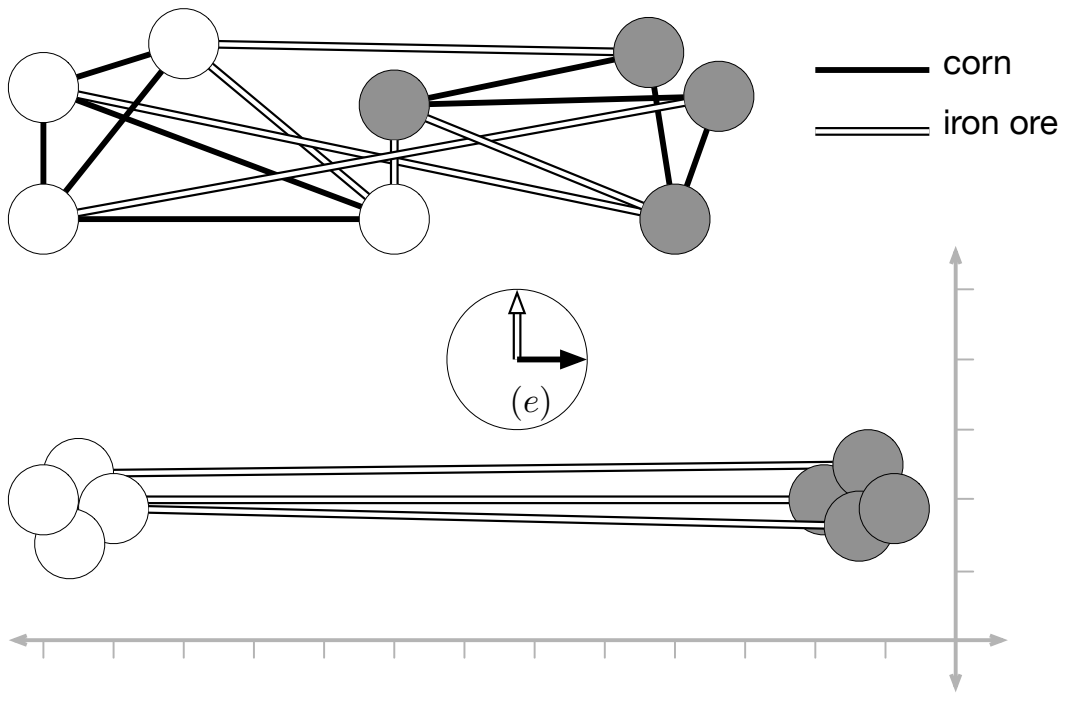

(b) If the white nodes are placed at the point $(-1, 0)$, and the gray nodes are placed at the point $(1, 0)$, the market experiences no tension, because the type vector for iron ore is orthogonal to the shock.

Figure 13: A shock on a muti-item network, showing the effect of choosing different type vectors.

products. Moreover, by the Johnson-Lindenstrauss lemma [JL84] one can always project the type vectors to a space whose dimension is proportional to $\log I$, while introducing tiny errors in the inner products. So, one can ensure that $d$ is much smaller than $n$.

A shock in the system is an $nd$ dimensional vector $\boldsymbol{x} \in \mathbb{R}^{nd}$, that can be interpreted as $n$ vectors of dimension $d$. The coordinates $\boldsymbol{x}_u \in \mathbb{R}^d$ correspond to the beliefs of node $u$ about the amounts of all items that are worth a unit of currency, and the inner product $\langle \boldsymbol{\tau}, \boldsymbol{x}_u \rangle$ corresponds to $u$'s belief about the value of the item whose type vector is $\boldsymbol{\tau}$.

Trade along an edge $e = \{u, v\}$ induces a normalizing force on $v$ of

$$w(e) \cdot \langle \boldsymbol{\tau}(e), \boldsymbol{x}_u - \boldsymbol{x}_v \rangle \cdot \boldsymbol{\tau}(e).$$

The energy of the edge is

$$(1/2) \cdot w(e) \cdot \langle \boldsymbol{\tau}(e), \boldsymbol{x}_u - \boldsymbol{x}_v \rangle^2 .$$

Thus, if $\boldsymbol{x}_u - \boldsymbol{x}_v$ is orthogonal to the type vector, then no force is induced and the energy is 0, since the parties agree on their beliefs for the item they trade. Note that these quantities remain identical if the type vector is replaced with its negation.

Given a shock $\boldsymbol{x}$, we define the tension of the shock to be the sum of the magnitudes of all forces felt by the nodes

$$\mathrm{T}(\boldsymbol{x}) = \sum_{u \in V} \left\| \sum_{\{u,v\} \in e \in E} w(e) \cdot \langle \boldsymbol{\tau}(e), \boldsymbol{x}_v - \boldsymbol{x}_u \rangle \cdot \boldsymbol{\tau}(e) \right\|,$$

where here $\|\boldsymbol{y}\| = \sqrt{\sum_{j=1}^{d} \boldsymbol{y}_j^2}$ denotes the length of the vector $\boldsymbol{y}$. Figure 13a shows the tension a network experiences when the shock is along a direction in one of the items, and the other item has a type vector that is not orthogonal. Figure 13b shows a similar situation when the type vectors are orthogonal. We define the energy of the shock to be the sum of the energies on all edges:

$$\mathrm{E}(\boldsymbol{x}) = (1/2) \sum_{\{u,v\} = e \in E} w(e) \cdot \langle \boldsymbol{\tau}(e), \boldsymbol{x}_u - \boldsymbol{x}_v \rangle^2 .$$

As in the single item case, the gradient of the energy is the amount of force needed at each node to hold the network in place:

$$(\nabla(\mathrm{E}(\boldsymbol{x})))_u = \sum_{u \in e = \{u,v\}} w(e) \cdot \langle \boldsymbol{\tau}(e), \boldsymbol{x}_u - \boldsymbol{x}_v \rangle \cdot \boldsymbol{\tau}(e),$$

So adjusting beliefs in the directions of the forces felt at each node is the quickest way to reduce the energy.

# 4 Rigorously Defining Market Connectivity

In this section, we give mathematical definitions for the tension and energy of trade networks and show how to identify bottlenecks mathematically. We aim to give numbers that capture the behavior of the network for all significant shocks. To ease comprehension, we explain the definition for the single item case first, and then generalize the definitions to handle multiple items.

## 4.1 Single Item Markets

Key to the definition is what kinds of shocks should be considered significant. A small, completely disconnected part of the network can give rise to 0 tension/energy. Suppose the network consists of two disjoint sets $\mathsf{A}$, $\mathsf{B}$ that have no edges between and the sum of all weights of edges in $\mathsf{A}$ is $\epsilon$ and the sum of all weights in $\mathsf{B}$ is $w - \epsilon$. If $\epsilon$ is very small, and $\mathsf{B}$ is well connected, this should be considered an efficient market. But placing all the nodes of $\mathsf{A}$ at $w - \epsilon$ and all the nodes of $\mathsf{B}$ at $-\epsilon$ gives $\mathrm{E}(\boldsymbol{x}) = \mathrm{T}(\boldsymbol{x}) = 0$, even though $\sum_{u \in V} w(u) \cdot \boldsymbol{x}_u = 0$. Such shocks focus too much attention on small anomalies in the network and should not be considered significant.

Observe that the tension and energy of a shock remain the same if every entry of the shock is shifted by the same number. Namely, if $\mathbf{1}$ denotes the all 1's vector, then $\mathrm{E}(\boldsymbol{x}) = \mathrm{E}(\boldsymbol{x} + \gamma \mathbf{1})$ and $\mathrm{T}(\boldsymbol{x}) = \mathrm{T}(\boldsymbol{x} + \gamma \mathbf{1})$, for every number $\gamma$. Keeping this in mind, it will be convenient to shift each shock so that the average belief of the nodes (weighted by the volume of trade) is 0. For any shock $\boldsymbol{x}$, define

$$\overline{\boldsymbol{x}} = \boldsymbol{x} - \frac{\sum_{u \in V} w(u) \cdot \boldsymbol{x}_u}{w} \cdot \mathbf{1}.$$

$\overline{\boldsymbol{x}}$ is the shock obtained by shifting $\boldsymbol{x}$ so that the average shock value is 0:

$$\sum_{u \in V} w(u) \cdot \overline{\boldsymbol{x}}_u = \sum_{u \in V} w(u) \cdot \boldsymbol{x}_u - \sum_{u \in V} w(u) \cdot \boldsymbol{x}_u = 0.$$

For any node $u$, $\overline{\boldsymbol{x}}_u$ is a measure of how far node $u$'s belief is from the average belief in the network. In order to preclude shocks that put undue weight on small parts of the network, we define the set of significant shocks of magnitude $\alpha$, for $0 \leq \alpha \leq 1$, to be:

$$\mathcal{V}_\alpha = \left\{ \boldsymbol{x} : \text{ for every } v \in V, \ \sum_{u \in V} \frac{w(u) \cdot |\overline{\boldsymbol{x}}_u|}{w} \geq \alpha \cdot |\overline{\boldsymbol{x}}_v| \right\}.$$

Significant shocks are those where the magnitude of the shock is well spread among the nodes in the sense that the shock affects at least an $\alpha$ (weighted) fraction of the trade in the market. Our measure of connectivity should be scale invariant. One can always increase or decrease the energy and tension by multiplying the shock by a scalar. To ignore such scaling factors, we measure the ratio of the tension and energy to a normalization factor:

**Definition 3.** *The* $\alpha$-tension *of the trade network is*

$$\mathrm{T}_\alpha = \min_{\boldsymbol{x} \in \mathcal{V}_\alpha} \frac{\mathrm{T}(\boldsymbol{x})}{2 \sum_{u \in V} w(u) \cdot |\overline{\boldsymbol{x}}_u|}.$$

Next we define the energy[7] of the trade network. Significant shocks in this context are given by:

$$\mathcal{U}_\alpha = \left\{ \boldsymbol{x} : \text{ for every } v \in V, \ \sum_{u \in V} \frac{w(u) \cdot \overline{\boldsymbol{x}}_u^2}{w} \geq \alpha \cdot \overline{\boldsymbol{x}}_v^2 \right\}.$$

**Definition 4.** *The* $\alpha$-energy *of the trade network is*

$$\mathrm{E}_\alpha = \min_{\boldsymbol{x} \in \mathcal{U}_\alpha} \frac{\mathrm{E}(\boldsymbol{x})}{\sum_{u \in V} w(u) \cdot \overline{\boldsymbol{x}}_u^2}.$$

The differences in the choices of the normalization factors and the choices of significant shocks for tension and energy can be explained by the properties they ensure. The proofs of these facts are elementary, and can be found in Appendix A.

**Fact 5.** $1 \geq \mathrm{T}_\alpha \geq 0$ *and* $1 \geq \mathrm{E}_\alpha \geq 0$.

Both quantities are relatively large when the network is well connected:

[7]The energy of the network is closely related to the concept of *conductance* in spectral graph theory. The expression for the energy is the same as the expression for the conductance of the normalized Laplacian of a graph.

**Fact 6.** *For every $\alpha$, the fully connected network on $n$ nodes with unit edge weights has*

$$\mathrm{T}_\alpha = \frac{1}{2} + \frac{1}{2(n-1)} = \mathrm{E}_\alpha.$$

When the number of participants is $n = 2$, Fact 6 shows that $\mathrm{T}_\alpha = 1 = \mathrm{E}_\alpha$. Since the set of significant shocks only gets bigger as $\alpha$ gets smaller, we have:

**Fact 7.** *For every $\alpha \leq \beta$, $\mathrm{T}_\alpha \leq \mathrm{T}_\beta$, $\mathrm{E}_\alpha \leq \mathrm{E}_\beta$.*

Although the set of significant shocks for tension and energy are different, they are closely related:

**Lemma 8.** $\mathcal{U}_\alpha \subseteq \mathcal{V}_\alpha \subseteq \mathcal{U}_{\alpha^2}$.

Finally, we have:

**Fact 9.** $\mathrm{E}_\alpha = 0 \Rightarrow \mathrm{T}_\alpha = 0 \Rightarrow \mathrm{E}_{\alpha^2} = 0$.

Let $B$ be a subset of $k$ nodes. Define $\mathrm{T}^B(\boldsymbol{x})$ to be the tension of the shock after deleting the edges of the network that do not touch $B$.

**Definition 10.** *A $(k, \alpha, \beta)$* bottleneck for tension *is a set of $k$ nodes $B$ such that that there is a vector $\boldsymbol{x} \in \mathcal{V}_\alpha$ with $\mathrm{T}(\boldsymbol{x}) > 0$ and*

$$\frac{\mathrm{T}^B(\boldsymbol{x})}{\mathrm{T}(\boldsymbol{x})} \geq \beta.$$

Similarly, let $\mathrm{E}^B(\boldsymbol{x})$ denote the energy due to all the edges that touch the set $B$.

**Definition 11.** *A $(k, \alpha, \beta)$* bottleneck for energy *is a set of $k$ nodes $B$ such that there is a vector $\boldsymbol{x} \in \mathcal{U}_\alpha$ with $\mathrm{E}(\boldsymbol{x}) > 0$ and*

$$\frac{\mathrm{E}^B(\boldsymbol{x})}{\mathrm{E}(\boldsymbol{x})} \geq \beta.$$

A $(k, \alpha, \beta)$ bottleneck is a serious obstruction to efficiency when $k$ is small and $\alpha, \beta$ are relatively large. Intuitively, when the network has a $(k, \alpha, \beta)$ bottleneck, there is a small set of $k$ nodes $B$ that is responsible for an unusually large share $\beta$ of the tension/energy of the network when the network experience a large shock of magnitude $\alpha$.

## 4.2 Markets with Multiple Items

As in the single item case, we define:

$$w(u) = \sum_{e \ni u} w(e),$$

to be the total weight of all edges that touch $u$, and

$$w = \sum_{u \in V} w(u)$$

to be sum of all the weights of each node.

We need a technical lemma whose proof is deferred to Appendix A:

**Lemma 12.** *For every shock $\boldsymbol{x}$, there is a vector $\boldsymbol{z} \in \mathbb{R}^d$ such that*

$$\sum_{u \in e \in E} w(e) \langle \boldsymbol{\tau}(e), \boldsymbol{x}_u \rangle \cdot \boldsymbol{\tau}(e) = \sum_{u \in e \in E} w(e) \langle \boldsymbol{\tau}(e), \boldsymbol{z} \rangle \cdot \boldsymbol{\tau}(e).$$

Let $\boldsymbol{y}$ be the vector such that for every $u \in V$, $\boldsymbol{y}_u = \boldsymbol{z}$, and define

$$\overline{\boldsymbol{x}} = \boldsymbol{x} - \boldsymbol{y}.$$

The point of this definition is that the energy and tension of $\overline{\boldsymbol{x}}$ remain exactly the same as those of $\boldsymbol{x}$,

$$\mathrm{E}(\boldsymbol{x}) = \mathrm{E}(\overline{\boldsymbol{x}}), \mathrm{T}(\boldsymbol{x}) = \mathrm{T}(\overline{\boldsymbol{x}}),$$

but the weighted average belief of $\overline{\boldsymbol{x}}$ is 0:

$$\sum_{u \in e \in E} w(e) \cdot \langle \boldsymbol{\tau}(e), \overline{\boldsymbol{x}}_u \rangle \cdot \boldsymbol{\tau}(e) = 0,$$

and $\overline{\boldsymbol{x}}$ still captures the inefficiencies described in $\boldsymbol{x}$.

Towards defining the $\alpha$-tension, for each $\alpha > 0$, we define the set of significant shocks to be

$$\mathcal{V}_\alpha = \left\{ \boldsymbol{x} : \text{ for every } v \in V, \sum_{u \in e \in E} \frac{w(e) \cdot | \langle \boldsymbol{\tau}(e), \overline{\boldsymbol{x}}_u \rangle |}{w} \geq \alpha \cdot \|\overline{\boldsymbol{x}}_v\| \right\}.$$

**Definition 13.** *For $\alpha > 0$, the* $\alpha$-tension *of the trade network is*

$$\mathrm{T}_\alpha = \min_{\boldsymbol{x} \in \mathcal{V}_\alpha} \frac{\mathrm{T}(\boldsymbol{x})}{2 \cdot \sum_{u \in e \in E} w(e) \cdot | \langle \boldsymbol{\tau}(e), \overline{\boldsymbol{x}}_u \rangle |}.$$

For the energy, we define significant shocks with the set:

$$\mathcal{U}_\alpha = \left\{ \boldsymbol{x} : \text{ for every } v \in V, \sum_{u \in e \in E} \frac{w(e) \cdot \langle \boldsymbol{\tau}(e), \overline{\boldsymbol{x}}_u \rangle^2}{w} \geq \alpha \cdot \|\overline{\boldsymbol{x}}_v\|^2 \right\}.$$

**Definition 14.** *For $\alpha > 0$, the* $\alpha$-energy *of the trade network is*

$$\mathrm{E}_\alpha = \min_{\boldsymbol{x} \in \mathcal{U}_\alpha} \frac{\mathrm{E}(\boldsymbol{x})}{\sum_{u \in e \in E} w(e) \cdot \langle \boldsymbol{\tau}(e), \overline{\boldsymbol{x}}_u \rangle^2}.$$

In the case that $\boldsymbol{\tau}(e)$ is the same vector for every edge $e$, these definitions are identical to those defined for the single item case. To understand the definitions, we prove several facts whose proofs can be found in Appendix A. Continuing the analogy with the case of single item markets, we have:

**Fact 15.** $1 \geq \mathrm{T}_\alpha \geq 0$ *and* $1 \geq \mathrm{E}_\alpha \geq 0$.

**Fact 16.** *For any choice of type vectors, if the network has $n$ nodes such that every node trades every item with every other node at unit rate, we have $\mathrm{T}_\alpha \geq \frac{\alpha}{2}$, and $\mathrm{E}_\alpha = \frac{1}{2} + \frac{1}{2(n-1)}$.*

Note that when the number of participants is $n = 2$, Fact 16 shows that $\mathrm{E}_\alpha = 1$. For large $n$, the energy of this completely connected network is close to 1/2. To contrast, the tension can actually be small. In Appendix B, we give an example showing that one can have fully connected networks where the tension is close to $\alpha/2$.

Since the set of significant shocks only gets bigger as $\alpha$ gets smaller, we have:

**Fact 17.** *For every* $\alpha \leq \beta$*,* $\mathrm{T}_\alpha \leq \mathrm{T}_\beta$ *and* $\mathrm{E}_\alpha \leq \mathrm{E}_\beta$*.*

**Lemma 18.** $\mathcal{U}_\alpha \subseteq \mathcal{V}_\alpha \subseteq \mathcal{U}_{\alpha^2}$

Finally, we have:

**Fact 19.** $\mathrm{E}_\alpha = 0 \Rightarrow \mathrm{T}_\alpha = 0 \Rightarrow \mathrm{E}_{\alpha^2} = 0$.

Given the changes we have already made, the definitions that capture bottlenecks are identical to the corresponding definitions in the single item case. Define $\mathrm{T}^B(\boldsymbol{x})$ to be the tension of the shock after deleting the edges of the network that do not touch $B$.

**Definition 20.** *A* $(k, \alpha, \beta)$ bottleneck for tension *is a set of* $k$ *nodes* $B$ *such that that there is a vector* $\boldsymbol{x} \in \mathcal{V}_\alpha$ *with* $\mathrm{T}(\boldsymbol{x}) > 0$ *and*

$$\frac{\mathrm{T}^B(\boldsymbol{x})}{\mathrm{T}(\boldsymbol{x})} \geq \beta.$$

Similarly, let $\mathrm{E}^B(\boldsymbol{x})$ denote the energy due to all the edges that touch the set $B$.

**Definition 21.** *A* $(k, \alpha, \beta)$ bottleneck for energy *is a set of* $k$ *nodes* $B$ *such that there is a vector* $\boldsymbol{x} \in \mathcal{U}_\alpha$ *with* $\mathrm{E}(\boldsymbol{x}) > 0$ *and*

$$\frac{\mathrm{E}^B(\boldsymbol{x})}{\mathrm{E}(\boldsymbol{x})} \geq \beta.$$

A $(k, \alpha, \beta)$ bottleneck is a serious obstruction to efficiency when $k$ is small and $\alpha, \beta$ are relatively large. Intuitively, when the network has a $(k, \alpha, \beta)$ bottleneck, there is a small set of $k$ nodes $B$ that is responsible for an unusually large share $\beta$ of the tension/energy of the network when the network experience a large shock of magnitude $\alpha$.

# 5 Discussion and Future Work

The principal contribution of this work is the definition of the trade network and the concepts of tension, energy and bottlenecks, which allow for a theory of market efficiency that does not suffer from the joint-hypothesis problem. The trade network provides a bridge from observable data to concepts like market efficiency, without making any assumptions about the evolution or correctness of prices. The trade network encodes data about the information that is visible to market participants and the information participants are able to act on. Market efficiency relies on who knows what, rather than what *everyone* knows, and the concept of the trade network provides metrics about who knows what.

We believe that the mathematics in our theory is useful to measure efficiency because the single item case is similar to equations used to model analogous concepts in physics. Experiments have confirmed that these equations accurately describe electrical networks, networks of springs, and the diffusion of heat. That said, one can only know if the choices we have made in the formulas for tension, energy and bottlenecks are well-founded by experimenting with real data. Experiments can help to validate the model, or find issues that suggest refinements to the model. Gathering data to approximate the trade network would be extremely useful.

It would be very useful to find sampling based techniques to estimate the energy/tension of the network, and use them to compute the statistics we have defined in this work. Can we sample small amounts of data about the trade network and use it to compute the tension and energy, and find bottlenecks?

There are several ways to extend our model that we believe could be fruitful:

- One could study the evolution of the trade network over time. An efficient market should have a rapidly evolving trade network, with traders switching trading partners to take advantage of new information and better prices. If nodes do not switch trading partners in response to shocks, that is an inefficiency. One could look for a model that rigorously captures this kind of market efficiency using data.
- One could study the relationship between the rates of trade and prices. If nodes do not often trade with partners offering them the best price, the market is likely to be inefficient. Again, it remains open to rigorously define this kind of efficiency.

## 6 Acknowledgements

Thanks to Paul Beame, Morgan Dixon, Abe Friesen, Shayan Oveis Gharan, Kevin Miniter, Shay Moran, Alex Jaffe, Anna Karlin, Kayur Patel, Sivaramakrishnan Natarajan Ramamoorthy, Darcy Rao, Tim Roughgarden, Brad Wagenaar, Alex White and Amir Yehudayoff for useful comments. Thanks to Kayur Patel for many suggestions that improved the figures in this paper.

# A Technical Proofs

*Proof of Fact 5.* The tension is clearly non-negative. We have:

$$\begin{aligned} \mathrm{T}(\boldsymbol{x}) = \mathrm{T}(\overline{\boldsymbol{x}}) &= \sum_{u \in V} \left| \sum_{v \in V} w(\{u,v\}) \cdot (\overline{\boldsymbol{x}}_u - \overline{\boldsymbol{x}}_v) \right| \\ &\leq \sum_{u \in V} \sum_{v \in V} w(\{u,v\}) \cdot (|\overline{\boldsymbol{x}}_u| + |\overline{\boldsymbol{x}}_v|) \\ &= 2 \sum_{u \in V} w(u) \cdot |\overline{\boldsymbol{x}}_u|, \end{aligned}$$

proving that $\mathrm{T}_\alpha \leq 1$, for $\boldsymbol{x} \in \mathcal{V}_\alpha$, since such $\boldsymbol{x}$ must have $2\sum_{u \in V} w(u) \cdot |\overline{\boldsymbol{x}}_u| > 0$.

Similarly, the energy is clearly non-negative. We have

$$\begin{aligned}
\mathrm{E}(\boldsymbol{x}) = \mathrm{E}(\overline{\boldsymbol{x}}) &= (1/2) \sum_{\{u,v\}\in E} w(\{u,v\}) \cdot |\overline{\boldsymbol{x}}_u - \overline{\boldsymbol{x}}_v|^2 \\
&\leq \sum_{\{u,v\}\in E} w(\{u,v\}) \cdot (|\overline{\boldsymbol{x}}_u|^2 + |\overline{\boldsymbol{x}}_v|^2) \qquad \text{since } 2a^2 + 2b^2 \geq (a+b)^2 \text{ for all } a, b. \\
&= \sum_{u\in V} w(u) \cdot |\overline{\boldsymbol{x}}_u|^2,
\end{aligned}$$

proving that $\mathrm{E}_\alpha \leq 1$, for $\boldsymbol{x} \in \mathcal{U}_\alpha$. ☐

*Proof of Fact 6.* For any shock $\boldsymbol{x} \in \mathcal{V}_\alpha$, we have

$$\begin{aligned}
\mathrm{T}(\boldsymbol{x}) = \mathrm{T}(\overline{\boldsymbol{x}}) &= \sum_{u\in V} \left| \sum_{v\in V} (\overline{\boldsymbol{x}}_u - \overline{\boldsymbol{x}}_v) \right| \\
&= \sum_{u\in V} \left| n \cdot \overline{\boldsymbol{x}}_u - \sum_{v\in V} \overline{\boldsymbol{x}}_v \right| \\
&= \sum_{u\in V} n \cdot |\overline{\boldsymbol{x}}_u|. \qquad \text{by the definition of } \overline{\boldsymbol{x}}
\end{aligned}$$

Thus we have

$$\mathrm{T}_\alpha = \min_{\boldsymbol{x}\in V_\alpha} \frac{\mathrm{T}(\boldsymbol{x})}{2\sum_{u\in V}(n-1)|\overline{\boldsymbol{x}}_u|} = \frac{n}{2(n-1)} = \frac{1}{2} + \frac{1}{2(n-1)}.$$

For any $\boldsymbol{x} \in \mathcal{U}_\alpha$, we have

$$\begin{aligned}
\mathrm{E}(\boldsymbol{x}) = \mathrm{E}(\overline{\boldsymbol{x}}) &= (1/2) \sum_{\{u,v\}\in E} (\overline{\boldsymbol{x}}_u - \overline{\boldsymbol{x}}_v)^2 \\
&= (1/2) \sum_{\{u,v\}\in E} \overline{\boldsymbol{x}}_u^2 + \overline{\boldsymbol{x}}_v^2 - 2\overline{\boldsymbol{x}}_u\overline{\boldsymbol{x}}_v \\
&= (1/2) \sum_{u\in V} (n-1)\overline{\boldsymbol{x}}_u^2 - 2 \sum_{\{u,v\}\in E} \overline{\boldsymbol{x}}_u\overline{\boldsymbol{x}}_v.
\end{aligned}$$

Now since $\sum_u \overline{\boldsymbol{x}}_u = 0$, we have that for every $u$, $\sum_{u\neq v} \overline{\boldsymbol{x}}_v = -\overline{\boldsymbol{x}}_u$. So

$$\begin{aligned}
\mathrm{E}(\boldsymbol{x}) &= (1/2) \sum_{u\in V} (n-1)\overline{\boldsymbol{x}}_u^2 - 2 \sum_{u\in V} \overline{\boldsymbol{x}}_u(-\overline{\boldsymbol{x}}_u) \\
&= (1/2) \sum_{u\in V} (n-1)\overline{\boldsymbol{x}}_u^2 + 2 \sum_{u\in V} \overline{\boldsymbol{x}}_u^2 \\
&= (1/2) \sum_{u\in V} (n+1)\overline{\boldsymbol{x}}_u^2.
\end{aligned}$$

Thus

$$\mathrm{E}_\alpha = \min_{\boldsymbol{x}\in\mathcal{U}_\alpha} \frac{\sum_{u\in V}(n+1)\overline{\boldsymbol{x}}_u^2}{2\sum_{u\in V}(n-1)\overline{\boldsymbol{x}}_u^2} = \frac{n+1}{2(n-1)} = \frac{1}{2} + \frac{2}{2(n-1)}.$$

☐

*Proof of Lemma 8.* Suppose $\boldsymbol{x} \in \mathcal{V}_\alpha$. Then we have

$$\begin{aligned}
\alpha \cdot |\overline{\boldsymbol{x}}_v| &\le \sum_{u \in V} \frac{w(u) \cdot |\overline{\boldsymbol{x}}_u|}{w} \\
&\le \sqrt{\sum_{u \in V} \frac{w(u)}{w}} \cdot \sqrt{\sum_{u \in V} \frac{w(u)}{w} \cdot \overline{\boldsymbol{x}}_u^2} && \text{by the Cauchy-Schwartz inequality} \\
&= \sqrt{\sum_{u \in V} \frac{w(u)}{w} \cdot \overline{\boldsymbol{x}}_u^2},
\end{aligned}$$

proving that

$$\sum_{u \in V} \frac{w(u) \cdot \overline{\boldsymbol{x}}_u^2}{w} \ge \alpha^2 \cdot \overline{\boldsymbol{x}}_v^2,$$

and so $\boldsymbol{x} \in \mathcal{U}_{\alpha^2}$. For the other containment, suppose that $\boldsymbol{x} \in \mathcal{U}_\alpha$, and $\overline{\boldsymbol{x}}_{v'}$ is the coordinate that has maximum magnitude. Then for every $v \in V$,

$$\begin{aligned}
\alpha \cdot |\overline{\boldsymbol{x}}_v| \le \alpha \cdot |\overline{\boldsymbol{x}}_{v'}| = \alpha \cdot \frac{|\overline{\boldsymbol{x}}_{v'}|^2}{|\overline{\boldsymbol{x}}_{v'}|} &\le \frac{1}{|\overline{\boldsymbol{x}}_{v'}|} \sum_{u \in V} \frac{w(u) \cdot \overline{\boldsymbol{x}}_u^2}{w} && \text{since } \overline{\boldsymbol{x}} \in \mathcal{U}_\alpha \\
&= |\overline{\boldsymbol{x}}_{v'}| \sum_{u \in V} \frac{w(u) \cdot \overline{\boldsymbol{x}}_u^2}{w \cdot \overline{\boldsymbol{x}}_{v'}^2} \\
&\le |\overline{\boldsymbol{x}}_{v'}| \sum_{u \in V} \frac{w(u) \cdot |\overline{\boldsymbol{x}}_u|}{w \cdot |\overline{\boldsymbol{x}}_{v'}|} && \text{since } \frac{|\overline{\boldsymbol{x}}_u|}{|\overline{\boldsymbol{x}}_{v'}|} \le 1 \\
&= \sum_{u \in V} \frac{w(u) \cdot |\overline{\boldsymbol{x}}_u|}{w},
\end{aligned}$$

so $\boldsymbol{x} \in \mathcal{V}_\alpha$. □

*Proof of Fact 9.* Suppose $\boldsymbol{x} \in \mathcal{U}_\alpha$ is a shock such that $\mathrm{E}(\boldsymbol{x}) = 0$. Then by lemma 8, $\boldsymbol{x} \in \mathcal{V}_\alpha$. Moreover, for every edge $\{u, v\}$ in the network, we have $\boldsymbol{x}_u = \boldsymbol{x}_v$, or else the energy would be positive. So we must have $\mathrm{T}(\boldsymbol{x}) = 0$, proving that $\mathrm{T}_\alpha = 0$.

Suppose $\boldsymbol{x} \in \mathcal{V}_\alpha$ and $\mathrm{T}(\boldsymbol{x}) = 0$. Then by lemma 8, $\boldsymbol{x} \in \mathcal{U}_{\alpha^2}$. Let $f(t) \in \mathbb{R}^n$ be the vector of forces felt by the nodes during the shock $t\boldsymbol{x}$. Then we see that $f(t)$ is $t$ times the vector of forces felt when the shock is $\boldsymbol{x}$, so $f(t) = \mathbf{0}$ for every $t$. Since this vector is the gradient of the energy at $t\boldsymbol{x}$, we must have $\mathrm{E}(1 \cdot \boldsymbol{x}) = \mathrm{E}(0 \cdot \boldsymbol{x}) = 0$. □

*Proof of Lemma 12.* This proof requires some knowledge about positive semidefinite matrices. View $\boldsymbol{\tau}(e)$ as a $d$ dimensional column vector, and consider the $d \times d$ matrix

$$A = \sum_{u \in e \in E} w(e) \cdot \boldsymbol{\tau}(e)\boldsymbol{\tau}(e)^\intercal.$$

We have

$$A\boldsymbol{z} = \sum_{u \in e \in E} w(e) \langle \boldsymbol{\tau}(e), \boldsymbol{z} \rangle \cdot \boldsymbol{\tau}(e),$$

and $A$ is positive semidefinite, since for every $\boldsymbol{z}$, we have

$$\boldsymbol{z}^\intercal A \boldsymbol{z} = \sum_{u \in e \in E} w(e) \cdot \langle \boldsymbol{\tau}(e), \boldsymbol{z} \rangle^2 \geq 0.$$

Thus there is an orthonormal basis $\boldsymbol{e}_1, \boldsymbol{e}_2, \ldots, \boldsymbol{e}_d$ for the space $\mathbb{R}^d$ and non-negative numbers $\gamma_1, \ldots, \gamma_d$ such that

$$A = \sum_{i=1}^{d} \gamma_i \cdot \boldsymbol{e}_i \boldsymbol{e}_i^\intercal.$$

Moreover, every type vector is contained in the span of vectors $\boldsymbol{e}_i$ for which $\gamma_i > 0$, because if, for example there is some edge for which $\boldsymbol{\tau}(e)$ is not contained in this span, then we can express $\boldsymbol{\tau}(e) = \boldsymbol{a} + \boldsymbol{b}$, where $\boldsymbol{a}$ is in the span, and $\boldsymbol{b} \neq \boldsymbol{0}$ is orthogonal to it. Then we have

$$0 = \boldsymbol{b}^\intercal \left( \sum_{i=1}^{d} \gamma_i \cdot \boldsymbol{e}_i \boldsymbol{e}_i^\intercal \right) \boldsymbol{b} = \boldsymbol{b}^\intercal A \boldsymbol{b} = \sum_{u \in e' \in E} w(e') \left\langle \boldsymbol{\tau}(e'), \boldsymbol{b} \right\rangle^2 \geq w(e) \left\langle \boldsymbol{\tau}(e), \boldsymbol{b} \right\rangle^2 = w(e) \cdot \|\boldsymbol{b}\|^2 > 0,$$

which is a contradiction. So, for $\boldsymbol{q} = \sum_{u \in e \in E} w(e) \cdot \langle \boldsymbol{\tau}(e), \boldsymbol{x}_u \rangle \cdot \boldsymbol{\tau}(e)$, we can set

$$\boldsymbol{z} = \sum_{i : \gamma_i > 0} (1/\gamma_i) \langle \boldsymbol{q}, \boldsymbol{e}_i \rangle \cdot \boldsymbol{e}_i,$$

and then we get

$$\sum_{u \in e \in E} w(e) \langle \boldsymbol{\tau}(e), \boldsymbol{z} \rangle \cdot \boldsymbol{\tau}(e) = A \boldsymbol{z} = \sum_{i=1}^{d} \langle \boldsymbol{q}, \boldsymbol{e}_i \rangle \cdot \boldsymbol{e}_i = \boldsymbol{q},$$

as required. □

*Proof of Fact 15.* The tension is clearly non-negative. We have:

$$\begin{aligned}
\mathrm{T}(\boldsymbol{x}) = \mathrm{T}(\overline{\boldsymbol{x}}) &= \sum_{u \in V} \left\| \sum_{\{u,v\} = e \in E} w(e) \cdot \langle \boldsymbol{\tau}(e), \overline{\boldsymbol{x}}_v - \overline{\boldsymbol{x}}_u \rangle \cdot \boldsymbol{\tau}(e) \right\| \\
&\leq \sum_{u \in V} \sum_{\{u,v\} = e \in E} w(e) \cdot \| \langle \boldsymbol{\tau}(e), \overline{\boldsymbol{x}}_v - \overline{\boldsymbol{x}}_u \rangle \cdot \boldsymbol{\tau}(e) \| \\
&\leq \sum_{u \in V} \sum_{\{u,v\} = e \in E} w(e) \cdot (| \langle \boldsymbol{\tau}(e), \overline{\boldsymbol{x}}_v \rangle | + | \langle \boldsymbol{\tau}(e), \overline{\boldsymbol{x}}_u \rangle |) \\
&= 2 \sum_{u \in e \in E} w(e) \cdot | \langle \boldsymbol{\tau}(e), \overline{\boldsymbol{x}}_u \rangle |,
\end{aligned}$$

proving that $\mathrm{T}_\alpha \leq 1$, for $\boldsymbol{x} \in \mathcal{V}_\alpha$.

Similarly, the energy is clearly non-negative. We have

$$\begin{aligned}
\mathrm{E}(\boldsymbol{x}) = \mathrm{E}(\overline{\boldsymbol{x}}) = (1/2) &\sum_{\{u,v\} = e \in E} w(e) \cdot \langle \boldsymbol{\tau}(e), \overline{\boldsymbol{x}}_u - \overline{\boldsymbol{x}}_v \rangle^2 \\
&\leq \sum_{\{u,v\} = e \in E} w(e) \cdot (\langle \boldsymbol{\tau}(e), \overline{\boldsymbol{x}}_u \rangle^2 + \langle \boldsymbol{\tau}(e), \overline{\boldsymbol{x}}_v \rangle^2) \quad \text{since } a^2 + b^2 \geq (a+b)^2/2 \text{ for all } a, b. \\
&= \sum_{u \in e \in E} w(e) \cdot \langle \boldsymbol{\tau}(e), \overline{\boldsymbol{x}}_u \rangle^2,
\end{aligned}$$

proving that $\mathrm{E}_\alpha \leq 1$, for $\boldsymbol{x} \in \mathcal{U}_\alpha$. □

*Proof of Fact 16.* Let $I$ denote the set of items, and let $\boldsymbol{\tau}(i)$ denote the type vector of item $i$. Then for any shock $\boldsymbol{x} \in \mathcal{V}_\alpha$, we have

$$\begin{aligned}
\mathrm{T}(\boldsymbol{x}) = \mathrm{T}(\overline{\boldsymbol{x}}) &= \sum_{u \in V} \left\| \sum_{v \in V, i \in I} \langle \boldsymbol{\tau}(i), \overline{\boldsymbol{x}}_v - \overline{\boldsymbol{x}}_u \rangle \cdot \boldsymbol{\tau}(i) \right\| \\
&= \sum_{u \in V} \left\| (n-1) \sum_{i \in I} \langle \boldsymbol{\tau}(i), \overline{\boldsymbol{x}}_u \rangle \cdot \boldsymbol{\tau}(i) - \sum_{v \in V, i \in I} \langle \boldsymbol{\tau}(i), \overline{\boldsymbol{x}}_v \rangle \cdot \boldsymbol{\tau}(i) \right\| \\
&= \sum_{u \in V} \left\| (n-1) \sum_{i \in I} \langle \boldsymbol{\tau}(i), \overline{\boldsymbol{x}}_u \rangle \cdot \boldsymbol{\tau}(i) \right\|. && \text{by the definition of } \overline{\boldsymbol{x}}
\end{aligned}$$

To bound this expression, we use the Cauchy-Schwartz inequality twice:

$$\begin{aligned}
&(n-1) \sum_{u \in V} \left\| \sum_{i \in I} \langle \boldsymbol{\tau}(i), \overline{\boldsymbol{x}}_u \rangle \cdot \boldsymbol{\tau}(i) \right\| \\
&\geq (n-1) \sum_{u \in V} \frac{\left\langle \sum_{i \in I} \langle \boldsymbol{\tau}(i), \overline{\boldsymbol{x}}_u \rangle \cdot \boldsymbol{\tau}(i), \overline{\boldsymbol{x}}_u \right\rangle}{\|\overline{\boldsymbol{x}}_u\|} && \text{by Cauchy-Schwartz} \\
&\geq (n-1) \sum_{u \in V, i \in I} \frac{\langle \boldsymbol{\tau}(i), \overline{\boldsymbol{x}}_u \rangle^2}{\|\overline{\boldsymbol{x}}_u\|}.
\end{aligned}$$

Now since $\overline{\boldsymbol{x}} \in \mathcal{V}_\alpha$, we have

$$\begin{aligned}
&\geq (n-1) \sum_{u \in V, i \in I} \langle \boldsymbol{\tau}(i), \overline{\boldsymbol{x}}_u \rangle^2 \cdot \frac{\alpha n |I|}{\sum_{u \in V, i \in I} |\langle \boldsymbol{\tau}(i), \overline{\boldsymbol{x}}_u \rangle|} && \text{since } \overline{\boldsymbol{x}} \in \mathcal{V}_\alpha \\
&\geq (n-1) \frac{\left( \sum_{u \in V, i \in I} |\langle \boldsymbol{\tau}(i), \overline{\boldsymbol{x}}_u \rangle| \right)^2}{n|I|} \cdot \frac{\alpha n |I|}{\sum_{u \in V, i \in I} |\langle \boldsymbol{\tau}(i), \overline{\boldsymbol{x}}_u \rangle|} && \text{by Cauchy-Schwartz} \\
&= \alpha(n-1) \sum_{u \in V, i \in I} |\langle \boldsymbol{\tau}(i), \overline{\boldsymbol{x}}_u \rangle|.
\end{aligned}$$

So we get

$$\mathrm{T}_\alpha = \min_{\boldsymbol{x} \in V_\alpha} \frac{\mathrm{T}(\boldsymbol{x})}{2 \sum_{u \in V, i \in I} (n-1) |\langle \boldsymbol{\tau}(i), \overline{\boldsymbol{x}}_u \rangle|} \geq \frac{\alpha}{2}.$$

Similarly, we have that for any shock $\boldsymbol{x} \in \mathcal{U}_\alpha$,

$$\begin{aligned}
\mathrm{E}(\boldsymbol{x}) = \mathrm{E}(\overline{\boldsymbol{x}}) &= (1/2) \sum_{u \neq v, i \in I} \langle \boldsymbol{\tau}(i), \overline{\boldsymbol{x}}_u - \overline{\boldsymbol{x}}_v \rangle^2 \\
&= (1/2) \sum_{u \neq v, i \in I} \langle \boldsymbol{\tau}(i), \overline{\boldsymbol{x}}_u \rangle^2 + \langle \boldsymbol{\tau}(i), \overline{\boldsymbol{x}}_v \rangle^2 - 2 \langle \boldsymbol{\tau}(i), \overline{\boldsymbol{x}}_u \rangle \langle \boldsymbol{\tau}(i), \overline{\boldsymbol{x}}_v \rangle \\
&= (1/2) \sum_{u, i \in I} (n-1) \langle \boldsymbol{\tau}(i), \overline{\boldsymbol{x}}_u \rangle^2 - \sum_{u \in u, i \in I} \langle \boldsymbol{\tau}(i), \overline{\boldsymbol{x}}_u \rangle \cdot \left\langle \boldsymbol{\tau}(i), \sum_{v \neq u} \overline{\boldsymbol{x}}_v \right\rangle. \qquad (2)
\end{aligned}$$

Now by the definition of $\overline{\boldsymbol{x}}$,

$$\mathbf{0} = \sum_{u \in V, i \in I} \langle \boldsymbol{\tau}(i), \overline{\boldsymbol{x}}_u \rangle \, \boldsymbol{\tau}(i) = \sum_{i \in I} \left\langle \boldsymbol{\tau}(i), \sum_{u \in V} \overline{\boldsymbol{x}}_u \right\rangle \cdot \boldsymbol{\tau}(i),$$

so

$$\begin{aligned} 0 &= \left\langle \sum_{i \in I} \left\langle \boldsymbol{\tau}(i), \sum_{u \in V} \overline{\boldsymbol{x}}_u \right\rangle \cdot \boldsymbol{\tau}(i), \sum_{u \in V} \overline{\boldsymbol{x}}_u \right\rangle \\ &= \sum_{i \in I} \left\langle \boldsymbol{\tau}(i), \sum_{u \in V} \overline{\boldsymbol{x}}_u \right\rangle^2, \end{aligned}$$

proving that for every $i$, $\left\langle \boldsymbol{\tau}(i), \sum_{u \in V} \overline{\boldsymbol{x}}_u \right\rangle = 0$, and $\left\langle \boldsymbol{\tau}(i), \sum_{v \neq V} \overline{\boldsymbol{x}}_v \right\rangle = - \langle \boldsymbol{\tau}(i), \overline{\boldsymbol{x}}_u \rangle$ . Returning to (2), we have

$$\begin{aligned} \mathrm{E}(\overline{\boldsymbol{x}}) &= (1/2) \sum_{u, i \in I} (n-1) \langle \boldsymbol{\tau}(i), \overline{\boldsymbol{x}}_u \rangle^2 - \sum_{u \in u, i \in I} \langle \boldsymbol{\tau}(i), \overline{\boldsymbol{x}}_u \rangle \left\langle \boldsymbol{\tau}(i), \sum_{v \neq u} \overline{\boldsymbol{x}}_v \right\rangle \\ &= (1/2) \sum_{u, i \in I} (n-1) \langle \boldsymbol{\tau}(i), \overline{\boldsymbol{x}}_u \rangle^2 + \sum_{u \in u, i \in I} \langle \boldsymbol{\tau}(i), \overline{\boldsymbol{x}}_u \rangle^2 \\ &= (1/2) \sum_{u, i \in I} (n+1) \langle \boldsymbol{\tau}(i), \overline{\boldsymbol{x}}_u \rangle^2 . \end{aligned}$$

This proves that

$$\mathrm{E}_\alpha = \min_{\boldsymbol{x} \in \mathcal{U}_\alpha} \frac{\mathrm{E}(\boldsymbol{x})}{\sum_{u \in V, i \in I} (n-1) \langle \boldsymbol{\tau}(i), \overline{\boldsymbol{x}}_u \rangle^2} = \frac{n+1}{2(n-1)} = \frac{1}{2} + \frac{1}{2(n-1)}.$$

□

*Proof of Lemma 18.* If $\boldsymbol{x} \in \mathcal{U}_\alpha$, let $v'$ be the node for which $\|\overline{\boldsymbol{x}}_{v'}\|$ is maximized. Then for every $v \in V$,

$$\begin{aligned} \alpha \cdot \|\overline{\boldsymbol{x}}_v\| \leq \frac{\alpha \cdot \|\overline{\boldsymbol{x}}_{v'}\|^2}{\|\overline{\boldsymbol{x}}_{v'}\|} \leq \sum_{u \in e \in E} \frac{w(e) \cdot \langle \boldsymbol{\tau}(e), \overline{\boldsymbol{x}}_u \rangle^2}{w \cdot \|\overline{\boldsymbol{x}}_{v'}\|} &= \|\overline{\boldsymbol{x}}_{v'}\| \sum_{u \in e \in E} \frac{w(e) \cdot \langle \boldsymbol{\tau}(e), \overline{\boldsymbol{x}}_u \rangle^2}{w \cdot \|\overline{\boldsymbol{x}}_{v'}\|^2} \\ &\leq \|\overline{\boldsymbol{x}}_{v'}\| \sum_{u \in e \in E} \frac{w(e) \cdot | \langle \boldsymbol{\tau}(e), \overline{\boldsymbol{x}}_u \rangle |}{w \cdot \|\overline{\boldsymbol{x}}_{v'}\|} \\ &= \sum_{u \in e \in E} \frac{w(e) \cdot | \langle \boldsymbol{\tau}(e), \overline{\boldsymbol{x}}_u \rangle |}{w}, \end{aligned}$$

so $\boldsymbol{x} \in \mathcal{V}_\alpha$. If $\boldsymbol{x} \in \mathcal{V}_\alpha$, then for every $v \in V$,

$$
\begin{aligned}
\alpha \cdot \|\overline{\boldsymbol{x}}_v\| &\le \sum_{u \in e \in E} \frac{w(e) \cdot |\langle \boldsymbol{\tau}(e), \overline{\boldsymbol{x}}_u \rangle|}{w} \\
&\le \sqrt{\sum_{u \in e \in E} \frac{w(e)}{w}} \cdot \sqrt{\sum_{u \in e \in E} \frac{w(e) \cdot \langle \boldsymbol{\tau}(e), \overline{\boldsymbol{x}}_u \rangle^2}{w}} \qquad \text{by the Cauchy-Schwartz inequality} \\
&= \sqrt{\sum_{u \in e \in E} \frac{w(e) \cdot \langle \boldsymbol{\tau}(e), \overline{\boldsymbol{x}}_u \rangle^2}{w}},
\end{aligned}
$$

proving that $\boldsymbol{x} \in \mathcal{U}_{\alpha^2}$. □

*Proof of Fact 19.* Suppose $\boldsymbol{x} \in \mathcal{U}_\alpha$ is a shock such that $\mathrm{E}(\boldsymbol{x}) = 0$. Then by Lemma 18, $\boldsymbol{x} \in \mathcal{V}_\alpha$. Moreover, for every edge $e = \{u, v\}$ in the network, we have $\langle \boldsymbol{\tau}(e), \boldsymbol{x}_u - \boldsymbol{x}_v \rangle = 0$, or else the energy would be positive. So we must have $\mathrm{T}(\boldsymbol{x}) = 0$, proving that $\mathrm{T}_\alpha = 0$.

Suppose $\boldsymbol{x} \in \mathcal{V}_\alpha$ and $\mathrm{T}(\boldsymbol{x}) = 0$. Then by Lemma 18, $\boldsymbol{x} \in \mathcal{U}_{\alpha^2}$. Let $f(t) \in \mathbb{R}^n$ be the vector of forces felt by the nodes during the shock $t\boldsymbol{x}$. Then we see that $f(t)$ is $t$ times the vector of forces felt when the shock is $\boldsymbol{x}$, so $f(t) = \mathbf{0}$ for every $t$. Since this vector is the gradient of the energy at $t\boldsymbol{x}$, we must have $\mathrm{E}(1 \cdot \boldsymbol{x}) = \mathrm{E}(0 \cdot \boldsymbol{x}) = 0$. □

# B Example Matching Fact 16

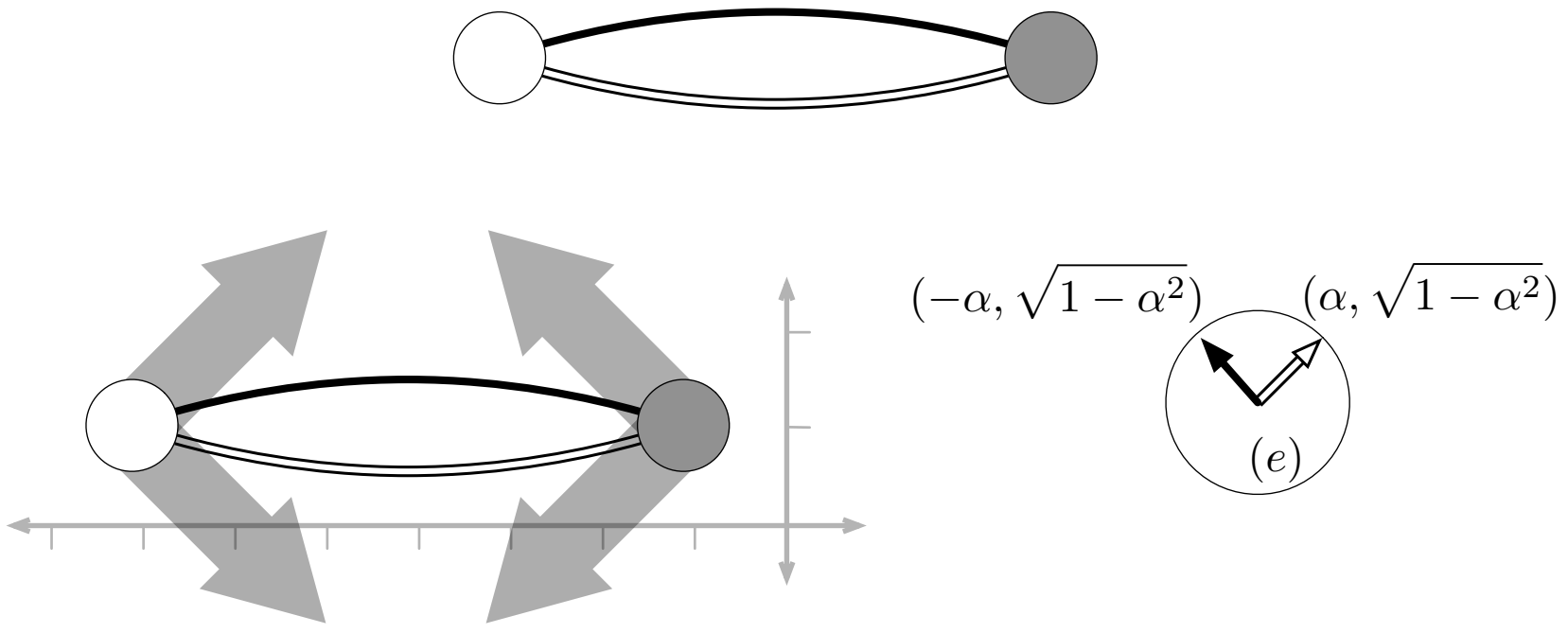

Figure 14: Even the fully connected network with 2 participants has $\mathrm{T}_\alpha = \alpha$.

Here we show that one can have a fully connected network whose tension is proportional to $\alpha$, matching the lower bound proved in Fact 16 upto a factor of 2. Similar ideas can be used to give a large network that asymptotically matches the bounds given by Fact 16.

Consider the network shown in Figure 14, which has just two participants, 1, 2. Suppose the first edge has type vector $\boldsymbol{\tau}(1) = (\alpha, \sqrt{1-\alpha^2})$, and the second has type vector $\boldsymbol{\tau}(2) = (-\alpha, \sqrt{1-\alpha^2})$, and both have weight 1. Then if we set $\boldsymbol{x}_1 = (1, 0)$ and $\boldsymbol{x}_2 = (-1, 0)$, we get that $\boldsymbol{x}$ is a significant shock with $\boldsymbol{x} = \overline{\boldsymbol{x}}$. Indeed, $\|\boldsymbol{x}_1\|, \|\boldsymbol{x}_2\| \leq 1$, and

$$\sum_{i=1,2} \sum_{u=1,2} |\langle \boldsymbol{\tau}(i), \boldsymbol{x}_u \rangle| = 4\alpha = \alpha w,$$

so $\boldsymbol{x} \in \mathcal{U}_\alpha$. But the tension of the shock is:

$$\begin{aligned}
\mathrm{T}(\boldsymbol{x}) &= \left\| \sum_{i=1,2} \langle \boldsymbol{\tau}(i), \boldsymbol{x}_1 - \boldsymbol{x}_2 \rangle \, \boldsymbol{\tau}(i) \right\| + \left\| \sum_{i=1,2} \langle \boldsymbol{\tau}(i), \boldsymbol{x}_2 - \boldsymbol{x}_1 \rangle \, \boldsymbol{\tau}(i) \right\| \\
&= 2 \left\| \sum_{i=1,2} \langle \boldsymbol{\tau}(i), \boldsymbol{x}_1 - \boldsymbol{x}_2 \rangle \, \boldsymbol{\tau}(i) \right\| \\
&= 2 \left\| 2\alpha \cdot (\alpha, \sqrt{1-\alpha^2}) - 2\alpha \cdot (-\alpha, \sqrt{1-\alpha^2}) \right\| = 8\alpha^2,
\end{aligned}$$

so we get $\mathrm{T}_\alpha \leq \frac{\mathrm{T}(x)}{2 \sum_{i=1,2} \sum_{u=1,2} |\langle \boldsymbol{\tau}(i), \boldsymbol{x}_u \rangle|} = \frac{8\alpha^2}{8\alpha} = \alpha$.